# THE COMPUTATIONAL PRIMITIVES OF ADAPTATION

**JONATHAN W. PAGE**[1]
PHARMACOLOGY AND PHARMACEUTICAL SCIENCES,
UNIVERSITY OF MISSOURI-KANSAS CITY
PREDAPTIC, LLC

ABSTRACT. Research on adaptive systems has traditionally focused on behavior (what organisms do) and mechanism (how their machinery works). This paper focuses on a third level, computation, which considers what adaptive systems must compute to survive and reproduce. It is proposed that adaptation has its own computational structure, comprising a small set of primitive operations common to all adaptive systems, regardless of their physical form. Six primitives—Arouse, Orient, Valence, Position, Boundary, and Attune—were selected using four criteria: necessity for existence, universality across independently evolved lineages, evolutionary conservation, and irreducibility. From this, two main implications follow. First, in biological systems, the primitives provide a substrate-neutral method for cross-species comparisons, reframing elaborative behaviors like attention, memory, and decision-making as combinations of these computations. Second, in artificial systems, the primitives offer a new way to view current challenges in machine intelligence, such as confabulation, prompt injection, distractibility, reward hacking, and catastrophic forgetting, suggesting that these issues may arise from a lack of these computations. Thus, for biological and artificial systems that persist, no specific physical substrate is necessary; rather, these primitives must be implemented if the systems are to be adaptive. What biology offers to inform machine intelligence, then, is not the brain's neural design but the computational functions it evolved to perform. The Computational Primitives Theory (CPT) presented here is a working hypothesis, intended for further refinement through discussion, empirical testing, and application.

## 1. INTRODUCTION

A honeybee discovers a patch of clover and forages for a while. As she flies back to the hive's dance floor, her nervous system encodes the three-dimensional flight path into a representation she shares with her nestmates through the waggle dance (von Frisch, 1967). Her sisters then decode the dance into directional and distance cues (Riley et al., 2005). Guided by this shared knowledge, they too are able to forage in the clover patch.

[1] Jonathan W. Page, ORCID: https://orcid.org/0009-0007-5637-0122. Please send correspondence to: jp6gp@umkc.edu (alternate: jonathan@predaptic.ai).

This progression, from navigation to communication to foraging, raises two questions that apply not only to bees but to all adaptive systems more broadly.

*Q1. What behaviors must an adaptive system exhibit to survive and propagate?* This, and questions like it, are generally answered by describing how organisms behave in their environments, evaluating the effectiveness of those behaviors, and comparing them across species.

*Q2. What mechanisms are necessary for an adaptive system to survive and propagate?* This question is often addressed by identifying the relevant neural and proto-neural machinery that gives rise to behavior and by tracing its circuitry to see how it contributes.

These first two questions address the functional and causal levels of adaptation, which have long been recognized in the study of behavior (Tinbergen, 1963). But there is a third question to ask when considering the honeybee's journey from flight to dance to foraging; one that addresses adaptive behavior more generally.

*Q3. What computations must an adaptive system perform to survive and propagate?* This question is not about behavior or neural machinery per se, but about the essential functions of adaptive systems that underlie both. Specifically, it asks about the computational processes that have evolved to meet the demands of survival and propagation.

This third question can be framed similarly to how Marr (1982) framed the levels of analysis in vision,[2] where he argued that the algorithmic and implementational levels cannot be fully understood apart from the computations they solve. The computational question posed in Q3 addresses adaptation and focuses not on the behavior an adaptive system produces (Q1) or the machinery that produces it (Q2), but on the problems any adaptive system must solve to persist, independent of the behaviors and mechanisms through which it solves them. Thus, the answer to Q3 does not lie in the bee's neural mechanisms that translate flight patterns nor in the behavior of her communicative dance; rather, it points to something more fundamental. It points to the computational structure of adaptive existence itself. By considering the bee from the Q3 perspective, the focus shifts to the operations that the bee, as an adaptive system, must perform to meet evolutionary imperatives, regardless of her underlying substrate, because adaptation itself has a level of analysis with its own characteristic problems to solve and operations to compute.

The term "computation," as used here and throughout, refers to how an organism performs perception and behavior, or, more generally, how a system processes inputs and produces outputs, as characterized at the information-processing level, independent of the physical machinery that performs them. A substrate that successfully produces the behavioral and mechanistic outcomes required for survival and reproduction, whether the bee or any other adaptive system, necessarily uses these computations. In other words, evolutionary pressure necessitates these computations, shaping the substrate over time to meet those demands.

The Computational Primitives Theory (CPT) proposed here formalizes the idea of a computational level of adaptation through a core set of substrate-neutral primitive operations (see Figure 1). By definition, these primitives must be specified at the computational level rather than the implementation level, and they must be present in any system that adapts, whether biological organisms, as discussed in the first part of this paper, or artificial systems, such as machine intelligence (if designed to be adaptive), as argued in the second part. Because these are computational rather than implementational requirements, the assertion is that any system that must maintain itself as an adaptive system under environmental pressure, whether carbon- or silicon-based, requires the same set of computations because it shares the same environment and information-structure space and faces the same demands. This narrower claim asserts that machines can be adaptive systems, too, if they are endowed with these computational primitives and placed in conditions that require them to adapt to persist.

[2] In his analysis of vision, Marr identified three levels for understanding information-processing systems: the computational level, which defines the problem being solved and why; the algorithmic level, which describes the procedures or representations used; and the implementation level, which identifies the physical mechanisms for carrying it out. Marr emphasized that the computational level is logically prior to the others, since you cannot fully understand a system without first specifying the problem it solves. While Marr focused on vision, the current framework applies this to the more general challenge of adaptive existence.

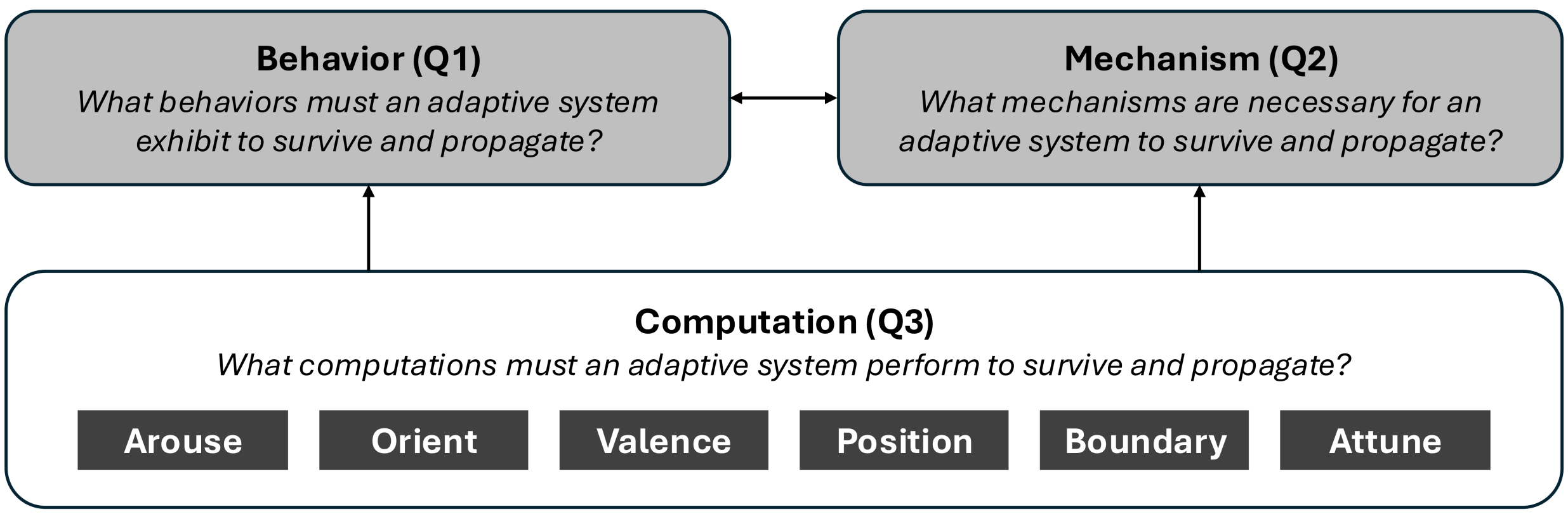

**Figure 1. Three levels of analysis for adaptive systems.** The behavior (Q1) and mechanism (Q2) questions represent two ways of thinking about adaptive systems and are addressed by established research traditions (their relationship is shown by a horizontal arrow). The computation question (Q3) concerns what an adaptive system must compute to survive and propagate, independent of the behavior it produces or the machinery that implements it. The Computational Primitives Theory (CPT) answers Q3 using six primitives (Arouse, Orient, Valence, Position, Boundary, Attune) that underlie both behavior and mechanism (as shown by the vertical arrows).

## 2. IN SEARCH OF THE COMPUTATIONAL PRIMITIVES

Identifying a core set of computational primitives is not straightforward. It raises difficult questions about what counts as a primitive. Without a clear framework or precise criteria, this task can easily devolve into merely cataloging the many useful functions found across different organisms. To avoid this, the "amoeba heuristic," articulated by R. Allen Gardner,[3] was used to create an initial list of candidate primitives. According to this heuristic, to determine whether a behavior is primary, ask yourself if an amoeba could do it. This is an informal way to categorize whether a function is primary or instead derived from advanced neural machinery or from a combination of more basic behaviors, including those shaped by neural mechanisms and species-specific processes unique to that organism.

[3] As a graduate student, I had the privilege of being in the same department as Allen and Beatrix Gardner at the University of Nevada, Reno. The Gardners, known for their cross-species research, such as teaching sign language to the chimpanzee Washoe (Gardner & Gardner, 1969), often discussed ways to separate primary from derived behaviors. Allen shared his amoeba heuristic informally in conversations and often in graduate seminars when discussing the origins of behavior.

Four criteria were formulated to identify primitive functions. They are proposed jointly as both necessary and sufficient for identifying a primitive function, although a strong interpretation of sufficiency (i.e., that the criteria generate a complete and unique set) is not assumed but left to empirical testing. Each criterion is first stated in a substrate-neutral form, which defines the requirement generally, and then in the biological diagnostic form used to detect it in organisms. Stating the criteria in this order shows that the requirement itself, not its biological application, is what generalizes beyond biology to all adaptive systems. Once the criteria were established, they were used to evaluate the candidate list of computations, from which the primitives included in this framework were selected:

a) *Necessity*. In substrate-neutral terms, the candidate computation must be essential to the system's continued existence as an adaptive system, such that its absence causes the system to fail. In biological terms, the computation must be essential to the organism's survival. Specifying that, to meet this criterion, an organism must persist underscores that the primitives are not merely optional features. Without them, a system could not persist. Grounding necessity in persistence rather than survival allows this criterion to be applied later to any non-biological adaptive system that must persist under pressure.
b) *Universality*. In substrate-neutral terms, the computation itself, not its physical form, must be present wherever adaptation occurs. The biological form is that the computation appears across a broad range of species and taxa, including lineages that diverged and evolved independently, and that it recurs regardless of implementation. The independent reappearance of a feature or function in lineages that could not have inherited it from a common ancestor shows that the feature is necessitated by the problem of adaptation itself rather than a product of any one substrate. If a function appears only in vertebrates, for example, it is a derived specialization rather than a primitive. Primitives should be observable wherever adaptation is necessary.
c) *Conservation*. In substrate-neutral terms, the candidate computation must be stably maintained wherever it appears, rather than arising only transiently. The biological form is that it must be implemented by evolutionarily conserved machinery, as evidenced by homologous circuitry in animals with nervous systems and by conserved signaling pathways and proto-cognitive machinery in simpler organisms. Preservation of the underlying machinery indicates that the function was selected for and maintained over evolutionary time. For clarity, universality (*b*) indicates that a feature recurs where it could not have been inherited; conservation shows that it is evolutionarily fundamental within the lineages that share it.[4]
d) *Irreducibility*. A primitive, by definition, is an operation that cannot be expressed as a combination of simpler operations from the same set. If a function can be described as a combination of two or more members of the set, it is considered derived, not primitive. Two qualifications prevent this criterion from becoming either circular or overreaching. First, because irreducibility is defined relative to the set, it establishes that the set is internally non-redundant. However, it does not, by itself, establish that the set is the exclusive decomposition of adaptive computation. Any alternative set of concepts could, in principle, be internally irreducible as well. What irreducibility means in this case is that the selected primitives constitute a sufficient and non-redundant set of computations, even though they may not be the only possible candidates. Second, showing that a behavior can be described in terms of the primitives is not the same as showing that it is constituted by them, since it is always possible to describe those same behaviors using a different set of words and descriptors. To show decomposition, the primitives must dissociate, demonstrating that they vary independently. Dissociation is therefore treated throughout as the empirical test of both whether a candidate is derived and whether two proposed primitives are distinct.

[4] It is important to note that conservation is about homology: the persistence of a feature or function through shared descent. Conservation alone shows that a feature is ancient, has been maintained by selection, and is retained in a particular lineage, but it does not necessarily imply that the feature is essential or universally required for adaptation. The necessity of a feature for adaptation is claimed by universality (criterion *b*). Thus, criteria *b* and *c* provide independent lines of evidence, where universality shows that a feature emerges even when it could not have been inherited, and conservation indicates its evolutionary significance within lineages that share it.

A candidate that fails any criterion cannot be classified as a computational primitive. For example, the "escape" behavior was initially considered a candidate. Although at first glance it satisfies the criteria of necessity, universality, and conservation, it does not meet the irreducibility standard. Escape draws on several basic processes, such as evaluating that something harmful is present (valence), locating the organism relative to the threat (position), directing attention to the route away from the threat (orientation), and mobilizing the energy to act (arousal). Because escape combines multiple processes and is therefore reducible to more basic ones, it cannot be considered computationally primitive under this framework. Additionally, the dissociation rule can be roughly applied to the basic processes underlying escape, showing that each of those operators can be selectively disrupted, moving them at least one level below escape: an organism may misevaluate a threat, mislocate it, fail to orient to an escape route, or fail to mobilize, any one of which can be disrupted while the others remain intact.

## 2.1 LANDING ON THE SIX

Based on the criteria outlined above, six candidates passed as computational primitives: Arouse, Orient, Valence, Position, Boundary, and Attune.[5] These were identified by applying the four criteria as filters to test each candidate and reject those that failed. However, a filter alone cannot establish that the surviving set is complete, because completeness ultimately depends on which candidates were considered in the first place. Thus, it may be informative to also approach this task from the opposite direction by briefly examining, independently of any candidate list, what adaptation minimally requires to ensure that at least the basics are included in the set.

An adaptive system faces a constantly changing, unpredictable environment that is indifferent to its persistence. In broad terms, only a few distinct demands must be met for a system to survive in these conditions. It must compute how much of its resources to allocate at any moment (Arouse), because sustained periods of high alertness or low energy can be fatal. Its computations must select, from a range of inputs, the features that matter most (Orient), as treating all inputs equally precludes predictive behavior. It must evaluate what it computes as favorable or unfavorable to its persistence (Valence), since selection without valuation cannot guide decisions or actions. It must computationally locate itself and other things within a shared spatial frame (Position), because action without spatial reference has unknown consequences. It must distinguish itself from its surroundings (Boundary), since a system that cannot computationally discriminate its own states and boundaries from the world's cannot protect or regulate itself. And it must adjust itself and its parameters based on what it encounters (Attune), because fixed behaviors and outputs cannot adapt to conditions that differ from those they were originally designed for or evolved to handle.

[5] These six are presented as one possible way to decompose adaptive systems. And because these six represent distinct computations, each should be selectively dissociable; that is, any one of them could be impaired, lost, or implemented independently. Thus, it is their dissociations, not their descriptions, that serve as the empirical test for distinguishing the primitives from simply relabeling similar behaviors.

Arriving at the same six primitives, both by filtering a candidate list and by considering the adaptive problem, is mutually reinforcing. Although it remains an open question whether these six form a complete set, the fact that two separate methods converge on the same conceptual terms increases confidence that this set captures the principal computational demands of adaptation. Together, these two approaches provide a strong basis for proposing these six as the full set.

Before concluding this section, it may be helpful to consider a few more candidates for the primitives list. These additional functions might reasonably be expected to be on the list; however, when the same logic used to exclude escape is applied to them, they are also excluded. (1) *Sensation* is the transduction of environmental energy into an internal signal. Orient and Valence presuppose this, since sensation is a prerequisite for computation rather than a computation that directly coordinates adaptive behavior. Even when treated as an operation, sensation does not dissociate from the primitives that use it. (2) *Action*, which involves producing motor output, serves as the means for the primitives to express themselves, rather than representing a separate adaptive computation. As previously discussed, escape shows how action-focused behavior can be decomposed. (3) *Reproduction* is necessary for lineages, so a species must reproduce to persist, but it is not a computation an individual must perform to persist. While the criteria involve survival and propagation, the six computational primitives are only meant to keep an individual organism alive long enough to reproduce. (4) *Anticipation*, (5) *communication*, and (6) *memory* were each considered and set aside as derived. *Anticipation* can be seen as an expression of Attune operating on Position and Valence; *communication* is a lineage- and channel-specific elaboration that fails to meet the requirement of universality; and storage-based *memory*, by dissociation, is inseparable from the experience-dependent parameter change brought about by Attune.

## 2.2 DESCRIBING THE PRIMITIVES

Below are detailed descriptions of each primitive, along with explanations of how each satisfies the four criteria established to test them (also, see Table 1).

### 2.2.1 Arouse

*General Description.* Arouse computes energetic readiness in organisms. When demands are placed on the system, the Arouse computation tracks and regulates the activation level needed to respond, modulating energetic readiness from quiescence to a highly responsive state.

*Necessity.* The ability to calibrate one's energetic state is crucial for survival. Without it, organisms must choose between staying in a constant state of high alert, which quickly depletes resources, or adopting a low-energy state to conserve resources, risking missed critical opportunities or failure to respond to threats. The Arouse computation addresses this dilemma by enabling the organism to dynamically match its readiness to the situation, conserving energy when possible and mobilizing it when necessary.

*Universality.* Energetic readiness is applied adaptively and flexibly in numerous ways. Amoebae, for example, alter their metabolic state when encountering nutrient gradients or chemical threats (Van Haastert & Devreotes, 2004; Gross & Pears, 2021). Fish exhibit an acute stress response when predators are near, including changes in activity level and oxygen intake (Wendelaar Bonga, 1997; Guo & Dixon, 2021). Insects show arousal-like behavioral changes in response to threats (Gibson et al., 2015; Kannan et al., 2022). Among mammals, researchers have documented changes in cortical gain, pupil dilation, and overall readiness (Joshi et al., 2016; Weiss et al., 2026), each modulated by task demands and environmental uncertainty.

**Table 1. The computational primitives in biological systems.** For each primitive, the computational function and the adaptive problem it solves are described, along with examples of its operation across four taxonomic categories. The Computational Primitives Theory (CPT) holds that these are the computations any adaptive system must perform to survive and reproduce. The breadth across taxa illustrates how the four criteria (necessity, universality, conservation, and irreducibility) favor substrate independence over species-specific elaboration.

| Primitive | Computation/ Adaptation | Example Taxonomies | | | |
|---|---|---|---|---|---|
| | | **Amoeba** | **Fish** | **Insect** | **Mammal** |
| ***Arouse*** | Adjusts energy levels to match current demands, so that activity is modulated to environmental conditions | Alter metabolic state according to nutrient availability or threat presence | Increase activity in response to predator signals | Adjust threat responses based on physiological state | Modulate arousal and readiness in response to salient or threatening cues |
| ***Orient*** | Selects sensory inputs by importance, so that limited processing resources are directed to relevant stimuli | Follow chemical gradients | Focus attention on prey or predator movement | Choose visual targets in flies and bees | Attention guided by stimulus salience |
| ***Valence*** | Evaluates whether to approach or avoid a stimulus, so that beneficial and harmful stimuli are distinguished | Move toward or away from chemical gradients | Approach food sources and avoid signs of predators | Learn through positive (appetitive) or negative (aversive) conditioning | Evaluate rewards and threats |
| ***Position*** | Determines one's location and orientation in space, so that position relative to the environment is established | Align body axis according to environmental gradients | Three-dimensional spatial orientation | Navigate using path integration and environmental landmarks | Form cognitive maps and encode specific locations |
| ***Boundary*** | Differentiates self from non-self, so that the organism's integrity as a distinct entity is preserved | Selective permeability across the cell boundary | Body schema formation and immune defense | Discriminate between self-generated and external stimuli | Body schema representation and self-recognition |
| ***Attune*** | Adjusts responses based on experience, so that the organism adapts to the conditions it encounters | Reduce responsiveness to repeated stimuli (habituation) | Learn through classical conditioning | Associative learning observed in bees and flies | Learning that integrates information across multiple sensory modalities |

*Conservation.* The machinery for regulating the energetic state is evolutionarily conserved. For example, single-celled organisms fine-tune their metabolic activity in response to environmental demands through complex signaling cascades (Wadhams & Armitage, 2004; Lyon, 2015; Jakowec & Finkel, 2025). In vertebrates, the locus coeruleus–norepinephrine system plays this role (Aston-Jones & Cohen, 2005; Sara, 2009; Hao et al., 2025), whereas invertebrates rely on functionally similar neuromodulatory systems, such as the octopamine network in arthropods (Roeder, 2005).

*Irreducibility.* No other primitive in the set can account for the essential regulation of the energetic state that Arouse provides.[6] Related behaviors, such as vigilance, alertness, and sleep-wake cycles, are best understood as expressions of Arouse rather than as distinct computational functions.

2.2.2 Orient

*General Description.* Orient is the computation that selects and prioritizes salience-weighted features for processing. It identifies which elements of an input require attention, given the organism's goals and state, and then allocates its limited processing resources to those that matter most.

*Necessity.* In any natural environment, the sheer volume of available information quickly exceeds what an organism can process. Yet, survival hinges on the computational ability to select what matters most. Treating all inputs equally would make it impossible to recognize or respond to the most crucial threats or opportunities. The Orient computation addresses this by prioritizing inputs, ensuring that processing resources are directed where they are most needed.

*Universality.* Selective orientation is a fundamental feature observed throughout biology. Amoebae, for example, move toward chemical signals indicating food and away from those indicating threat (Insall, 2010; Insall et al., 2022). Fish focus on specific motion patterns associated with prey or predators (Bianco & Engert, 2015; Temizer et al., 2015; Baier & Scott, 2024). Research on insects has revealed sophisticated selective attention, including the ability to filter competing visual targets, as seen in flies and bees (de Bivort & van Swinderen, 2016; Spaethe et al., 2006; Robert et al., 2024). Mammals also demonstrate selective attention, particularly in complex environments, with observable orienting responses to novel stimuli, movement, and conspecific cues across many species (cf. Knudsen, 2007; Poort et al., 2022; Krauzlis et al., 2023).

*Conservation.* The mechanisms for selective input are deeply conserved. For instance, the chemotaxis signaling pathway in bacteria and amoebae biases movement toward beneficial gradients (Swaney et al., 2010; Sourjik & Wingreen, 2012; Colin et al., 2021; Wheeler et al., 2024). Structures such as the superior colliculus and pulvinar in vertebrates (Krauzlis et al., 2013; Hafed et al., 2023) and the lobula in dragonflies and other flying insects (Wiederman & O'Carroll, 2013; Evans et al., 2022) perform attentional selection by weighting inputs based on their salience.

*Irreducibility.* Without the Orient computation, the system would have no way to prioritize competing inputs. None of the other primitives perform this calculation. Such a loss would completely undermine a system's ability to adapt. Behaviors related to this, such as target tracking and distractor suppression, are better understood as outcomes of the Orient computation under different conditions.

[6] Arouse is also proposed as allostasis (Sterling, 2012; Sterling & Laughlin, 2015), raising the question of whether energetic readiness is indeed primitive or simply an expression of a more general regulatory function. The universality and conservation evidence support both interpretations. The distinction may then depend on the irreducibility criterion. If energetic readiness dissociates from the experience-dependent regulation of Attune, for example, then Arouse would meet the criteria for being classified a primitive. If not, then the allostatic interpretation would be favored.

### 2.2.3 Valence

*General Description.* The Valence computation assigns approach-avoidance weights to inputs and internal states. Having a bipolar evaluative signal for features and states guides adaptive behavior toward beneficial options and away from harmful ones.

*Necessity.* To select an adaptive response, an organism must first evaluate the significance of what it encounters. It must determine whether something is positive (good/approach) or negative (bad/avoid). Without this ability, all signals would be processed as neutral, and behavior would become arbitrary with respect to its survival value. Valence solves this by generating a graded response that skews positive to encourage approach and negative to promote avoidance. This biases the organism's actions, guiding behavior in ways that are evolutionarily advantageous.

*Universality.* The approach-avoidance response is universal across taxa. Amoebae approach nutrient sources and avoid chemical irritants (Keizer-Gunnink et al., 2007; Kirolos & Gomer, 2022; El-Sobky et al., 2025). Fish approach food cues and avoid predator cues, with response times and intensities aligning with those behaviors (Domenici & Hale, 2019; Wee et al., 2022). Insects show classical conditioning to both positive and negative stimuli, demonstrating that valuation stimuli influence decision-making and learning (Menzel, 2012; Paoli et al., 2023). Many mammal species reliably approach palatable foods, conspecifics, and safe locations, and avoid predator odors, painful stimuli, and dangerous locations, while displaying affective facial and vocal expressions that match those behaviors (Anderson & Adolphs, 2014; Chen, 2022).

*Conservation.* Approach-avoidance behavior is evident across the evolutionary tree. Single-cell organisms exhibit rudimentary valence, with receptor systems that discriminate between beneficial and harmful chemical gradients and generate downstream signals that bias their movement accordingly (Hazelbauer et al., 2008; Guo et al., 2023). In vertebrates, valence is implemented through conserved circuits in the amygdala and ventral striatum (Janak & Tye, 2015; Smith & Torregrossa, 2021; Hagihara & Lüthi, 2024). Valence-evaluative circuits analogous to these are found in the mushroom bodies of insects (Aso et al., 2014; Modi et al., 2020).

*Irreducibility.* Any system that cannot compute basic Valence will not be able to respond adaptively to its environment. No other primitive provides such valuation. Many behaviors related to Valence, such as fear responses, appetitive seeking, and disgust responses, are better understood as being derived from it.

### 2.2.4 Position

*General Description.* The Position computation determines an organism's location and orientation relative to the environment and its own body. It provides a spatial reference frame in which all other primitives can operate. This allows the system to compute where it is in space, where other things are, and how it aligns with them.

*Necessity.* If an organism lacks a spatial frame of reference and cannot locate itself in space, it will be unable to navigate, approach, avoid, or coordinate its movement in a meaningful way. Actions without such spatial context have unknown consequences. The Position primitive solves this problem. It computes where the organism is and how it is oriented to guide its behavior in space.

*Universality.* The spatial computation of Position is evident across taxa. Amoebae use a polarized body axis to guide movement along chemical gradients (Iglesias & Devreotes, 2008; Kuhn et al., 2021). Fish use multiple sensory modalities to maintain three-dimensional spatial orientation (Salas et al., 2003; Baratti et al., 2022). To locate themselves in space, insects perform sophisticated spatial computations, including path integration and landmark-based navigation in bees, ants, and flies (Collett et al., 2013; Freas & Spetch, 2023). Mammals also use landmark and path-integration information to build cognitive maps for navigating complex environments (McNaughton et al., 2006; O'Keefe, 2025). Evidence of homing behavior, route learning, and spatial memory is documented across diverse mammalian lineages (Geva-Sagiv et al., 2015; Freas & Cheng, 2022).

*Conservation.* The computation of spatial reference in Position is conserved across evolution. Single-celled organisms maintain spatial reference through gradient sensing, which establishes front-back polarity along the body axis (Janetopoulos & Firtel, 2008; Ghose et al., 2022). In the mammalian hippocampal-entorhinal system, place cells and grid cells implement spatial reference (Moser et al., 2008; Dong & Fiete, 2024). Functionally analogous head-direction and place-coding circuits have been identified in invertebrates, including flies, bees, and ants (Seelig & Jayaraman, 2015; Stone et al., 2017; Hulse et al., 2021).

*Irreducibility.* A system without Position cannot organize any spatially directed behavior, making most adaptive behavior practically impossible. Many behaviors that appear to compute something similar to Position and therefore seem primary, such as navigation, homing, and postural control, are simply expressions of Position, since Position is dissociable from them.

2.2.5 Boundary

*General Description.* Boundary maintains an organism's integrity, creating the sense of a distinct entity. It is the computation that distinguishes the self from the non-self. It monitors permeability, tracking the organism's boundaries and what crosses them. It regulates what is allowed in and what is kept out, and identifies what belongs to the organism.

*Necessity.* An organism that cannot tell itself apart from the environment cannot modulate itself and the environment in ways that sustain its own existence. Without the Boundary computation, an organism cannot distinguish its own signals from external ones. The result is an inability to identify threats to its integrity and, therefore, to protect itself. Boundary solves this by computing and maintaining the self-other distinction.

*Universality.* The self-other property of Boundary is observable across taxa. Amoebae maintain cellular boundaries that distinguish their cytoplasm from the external environment and regulate bidirectional transport across the membrane (Maniak, 2002; Williams, 2010; Kay, 2021). Fish pass mirror and body-schema self-recognition tests (Kohda et al., 2019; Kohda et al., 2023; Kobayashi et al., 2024). Insects show evidence of self-representation in motor control and grooming behaviors that depend on distinguishing self-generated from environmental stimuli (Seeds et al., 2014; Webb, 2004; Zhang et al., 2020). Mammals show robust self-representation, including body schema, kinesthetic awareness, and discrimination between self-generated and externally caused sensations (Crapse & Sommer, 2008; Lei, 2023). Some primates (Gallup, 1970), cetaceans (Reiss & Marino, 2001), and elephants (Plotnik et al., 2006) have shown impressive levels of self-recognition.

*Conservation.* This self-other machinery is evident across evolution. Single-celled organisms use the selective permeability of their cell membranes to maintain boundaries (Nicolson, 2014; Nicolson & Ferreira de Mattos, 2022). Their molecular machinery distinguishes self from non-self at the chemical level, which underlies immune recognition in multicellular organisms (Janeway & Medzhitov, 2002; de Oliveira Mann & Hornung, 2021). Body-schema circuits in the parietal cortex of vertebrates compute something similar (Maravita & Iriki, 2004; Castro et al., 2023), providing a self-other distinction. Invertebrates have analogous body-representation circuits that implement self-representation (Poulet & Hedwig, 2007; Daly & Dacks, 2023).

*Irreducibility.* While other primitives seem to share a definition with Boundary, Boundary is argued to have a distinct computation. For example, Position locates an organism in space but cannot specify what counts as the organism. Orient selects which features to prioritize for processing, but doing so presupposes an organism that is doing the selecting. Immune responses, self-grooming, and defensive posturing are boundary-type behaviors, but these are better understood as expressions of Boundary. Although Boundary spans phenomena that appear disparate, such as membrane integrity, immune self/non-self recognition, body schema, and the discrimination of self-generated from externally generated signals, the Computational Primitives Theory (CPT) treats them as a single computation because they all address the same problem of determining what belongs to the organism. An organism that is unable to compute Boundary has no way to identify or protect itself, making it incapable of persisting.

#### 2.2.6 Attune

*General Description.* Attune modifies an organism's internal processes and parameters based on repeated, expected patterns in both the environment and its internal activity. It calibrates the organism as needed to operate effectively across varying conditions by synchronizing with the environment through tuning sensitivity, response thresholds, and predictive patterns.

*Necessity.* An organism whose parameters are fixed at birth cannot adapt to conditions that differ from those its ancestors encountered. Because environments vary in space and time, the conditions any organism faces will not be identical to those in which its species evolved, so it will encounter new problems to solve. Attune addresses these novel events by allowing the organism to update its operating parameters based on experience. This internal-to-external coherence, or synchronization, provides a fit that genetics alone cannot afford.

*Universality.* Experience-dependent coherence is evident across taxa. Amoeboid slime molds and other non-neural organisms exhibit habituation to repeated stimuli, reducing their response over time (Boisseau et al., 2016; Vogel & Dussutour, 2016; Boussard et al., 2021). Fish show classical conditioning, place learning, and habituation across multiple stimulus modalities (Bshary & Brown, 2014; Salena et al., 2021). Insects demonstrate robust learning, with conditioned responses and habituation documented in bees and flies (Perry et al., 2013; Finke et al., 2023; Rivi et al., 2023). Mammals exhibit extensive experience-dependent learning across modalities, as evidenced by classical conditioning, operant learning, perceptual learning, and habituation across mammalian taxa (Pearce & Bouton, 2001; Bouton et al., 2021).

*Conservation.* The ability to adjust parameters in an experience-dependent manner is evident across evolution. Single-celled organisms exhibit experience-dependent changes in receptor sensitivity and response thresholds (Dussutour, 2021). In nervous systems, mechanisms of synaptic plasticity, such as Hebbian learning, long-term potentiation, and spike-timing-dependent plasticity, implement the Attune computation (Malenka & Bear, 2004; Caporale & Dan, 2008; Debanne & Inglebert, 2023; Matsumoto et al., 2024). Analogous plasticity mechanisms have been observed in invertebrate nervous systems, including mollusks and arthropods (Kandel, 2001; Glanzman, 2010).

*Irreducibility.* A system that lacks the Attune computation cannot adapt to novel conditions, limiting its adaptive functions to whatever its genetics happen to fit. No other primitive can fulfill this need for adapting. Many behaviors that appear primary, such as habituation, classical and operant conditioning, and perceptual learning, can be understood as expressions of the Attune primitive.

Taken together, these six primitives are proposed to answer Q3 because they are the computations adaptive systems must perform to persist, and in doing so, to survive and propagate. Each was tested against the same four criteria of necessity, universality, conservation, and irreducibility. The extent to which any candidate satisfies these criteria is an empirical question; however, the converging evidence presented above suggests that these six indeed meet them. When evidence from the literature lends support, it indicates that a particular computation is present because the challenge of adaptation requires it, not simply because one substrate happens to provide it. In other words, when the same primitive appears across multiple independently evolved lineages, it provides stronger evidence that the computation is present in response to the universal demands of adaptation rather than being part of any one mechanism. Thus, these six should not be seen as what any particular system does, but as what any adaptive system needs to do to exist.

### 2.3 RELATIONSHIP TO EXISTING FRAMEWORKS

The idea that behavior derives from a limited set of primary functions is not new. Existing theories, such as active inference, enactivism, and basal cognition, already describe the underlying causes of adaptive behavior. Hence, it may be informative to situate CPT within these traditions. The purpose here is not to describe or even summarize each theory but to identify one or two areas within each where CPT most closely aligns.

Perhaps the closest comparison is with the free-energy principle and active inference (Friston, 2010; Friston et al., 2017), which explain adaptive functions top-down from a single variational objective. The Markov blanket (Kirchhoff et al., 2018) is somewhat similar to Boundary, though the blanket concerns statistical relationships while Boundary refers to a classification process. And precision-weighting (Feldman & Friston, 2010) is conceptually similar to Orient. CPT differs in that it is built from comparative criteria in a bottom-up manner, with the dissociations of the primitives defined by criteria independent of any stipulated generative model. Whether the six primitives in CPT should be included in the set becomes an empirical question rather than a modeling decision.

The autopoietic tradition (Maturana & Varela, 1980), extended by enactivists who ground value in precarious self-maintenance (Weber & Varela, 2002; Di Paolo, 2005), aligns with CPT’s insistence that viability is necessary. Boundary defines the self/non-self distinction that autopoiesis grounds in material self-production as a classification computation, and Valence functionalizes the claim that value requires a self with something at stake. In both cases, CPT treats the computation as sufficient rather than requiring material self-production, which matters for later discussions of a machine’s operational viability.

Basal cognition provides much of the comparative evidence for the universality and conservation criteria used to define the primitives, and its illustrative toolkit of capacities serves as a precedent for enumeration (Lyon et al., 2021; Levin, 2022). CPT proposes a closed, irreducible set of computations with a specified interaction structure. And the primitives are grounded in what adaptation requires rather than in what qualifies as cognition.

Situated alongside these theories, CPT draws on the comparative breadth of basal cognition, follows the enactive grounding of value in precarious self-maintenance, and operationalizes the free-energy principle to offer an explicitly enumerated and individually falsifiable set of computations necessary for adaptation. The next two sections further develop this contribution by treating the primitives as necessary for adaptive existence in two domains. In biology, the primitives reframe what cross-species comparison studies can coherently ask. In artificial systems, the primitives are turned outward as a diagnostic set to reveal what is missing from current machine intelligence and to show what must change to achieve a more general adaptive intelligence.

## 3. IMPLICATIONS OF THE SIX PRIMITIVES FOR BIOLOGICAL SYSTEMS

With the set of primitives enumerated, the discussion now turns to the framework’s practical significance. While the previous sections established the primitives, this section examines what the primitive framework, taken as a whole, offers for comparative research. Specifically, it asks: If the primitives represent the computations necessary for adaptive existence and apply across the evolutionary tree, what benefits could they provide for recognizing and characterizing core biological cognition?

Beyond providing the computational structure of adaptive existence, the primitives offer a level of analysis that current comparative research often lacks. That is, the primitives enable principled, substrate-neutral cross-species comparisons that sidestep the need to translate findings ad hoc (from human or rodent paradigms, for instance). Approaching comparative research from the perspective of the primitives makes several challenges in this field more tractable (cf. Bräuer et al., 2020). For example, rather than asking whether a given species possesses attention, memory, or self-recognition—questions often constrained by species-specific operationalizations (Schubiger et al., 2020)—it can instead be asked how that species' primitives function and interact. This reframing creates opportunities to ask and address new research questions that elaborative-behavior frameworks struggle to articulate.

### 3.1 THE PRIMITIVES AS SUBSTRATE-NEUTRAL CONCEPTS

A common problem in comparative research is inconsistent measurement across species. Protocols developed to measure a specific process in one species can rarely be used to measure it in another without modification (cf. Alessandroni et al., 2024). Take attention as an example. It has been studied in primates using eye-tracking and reaction-time paradigms (Hopper et al., 2021), but no such equivalents exist for bees or octopuses. Memory is another example. Researchers often operationalize it using human episodic-memory tasks (Davies & Clayton, 2024), but these tasks do not translate cleanly to species whose perception of time and space may differ fundamentally. These issues are common and nearly always require modifications to both test design and the interpretation of results.

The six primitives proposed in this paper offer an alternative. In fact, the four criteria used to identify the primitives (necessity, universality, conservation, and irreducibility) were chosen precisely to select functions that operate across the evolutionary tree. Given the way the criteria were constructed, they do not yield primitive concepts that merely happen to apply across species; they yield concepts selected to apply to any adaptive system. Cross-substrate generality, in other words, is what the criteria test for.

The methodological consequence is straightforward. For cross-species comparisons to be coherent, the concepts being compared must be defined at a level of analysis where such comparisons are meaningful. The six primitives offer a candidate set because, at the computational level of analysis, concepts can be clearly defined and applied across species. Thus, CPT's contribution to comparative research is that the primitives provide a level of analysis at the irreducible foundation of computation that underlies all adaptive systems. Further, CPT supports cross-species comparisons using principled criteria rather than ad hoc translations.

### 3.2 REFRAMING CONCEPTS AT THE PRIMITIVES LEVEL

To evaluate the usefulness of CPT in this regard, it may be informative to select a few widely studied cognitive processes and see how they can be explained using the primitives and their combinations. The following five examples show that cross-species tests are most informative at the level of the computational primitives, and that once a cross-species connection is established at this level, more species-specific functions can be understood as elaborations of behaviors derived from the primitives. The elaborations will vary across species depending on factors such as a species' computational machinery, its characteristic behaviors, and its environment—the very features that make comparative research challenging to begin with—so it is important to note that they all begin with the primitives as the foundation.

#### 3.2.1 Attention

As commonly studied, attention aligns best with the Orient primitive modulated by Arouse. The various forms of attention identified in the experimental literature, such as bottom-up versus top-down, sustained, selective, and divided (Lindsay, 2020), may simply be different instances of the same underlying Orient computation, modulated by varying Arouse states. Thus, the species-comparable question is not "Does this organism have attention?" with its accompanying species-specific operationalizations, but rather "Does this organism exhibit the Orient computation, and is Orient modulated by the Arouse computation?" The computationally-framed version of this question is answerable in amoebae (chemotactic selection modulated by metabolic state), in bees (selective visual response modulated by foraging state), and in mammals (attentional selection modulated by arousal state). Using a combination of the primitives Orient and Arouse makes cross-species comparisons of attention more coherent.

#### 3.2.2 Memory

Defining memory as experience-dependent behavior change aligns closely with the Attune primitive. When memory is considered in humans and rodents (cf. Sherman et al., 2023), pairing Attune with other primitives makes memory easier to test in other species. For example, Attune can be coupled with Position for spatial memory, with Boundary for self-referential memory, and so forth. In this light, how Attune operates in the species under study can drive the measurement question rather than whether that species has "memory" as measured in studies of human or rodent cognition. Habituation in amoebae, classical conditioning in fish, place learning in bees, and skill acquisition in mammals all demonstrate how Attune operates across substrates. The narrower question of whether a species has memory similar to humans dissolves into the more tractable question of how Attune pairs with the other primitives in each case.

#### 3.2.3 Decision-Making

Using the primitives, decision-making can be framed as Valence coordinating with Arouse and Orient. For example, an organism attends to the available options (Orient), then evaluates each for approach-avoidance valuation (Valence), and modulates those primitives according to its current readiness state (Arouse). Starting from the primitives, questions in cross-species decision research can focus on how the primitives are coupled within the species under study rather than on whether that species can make decisions in the sense of deliberative cognition. Foraging choices in bees, prey selection in fish, predator avoidance in amoebae, and complex choice behavior in mammals are simply expressions of the same primitive combinations operating to varying degrees of elaboration (Mobbs et al., 2018; Grima et al., 2025).

#### 3.2.4 Spatial Cognition

Spatial cognition can be viewed as the Position primitive working alongside Orient and Attune. In other words, an organism's capacity to locate itself in the environment, navigate through it, and remember spatial relationships over time is the joint work of at least these three primitives. Cross-species research on spatial cognition then becomes a question of how Position is implemented in the species under study, i.e., through gradient sensing in amoebae, path integration in bees, cognitive maps in mammals, etc. (Jeffery et al., 2024; cf. Kuhn et al., 2021). This reframes the question of whether a species has spatial cognition into the more operationalized question of how Position is implemented. According to CPT, all adaptive systems have a Position computation, but Position couples with other primitives in ways that vary across species to support increasingly elaborate spatial behavior.

### 3.2.5 Self-Other Discrimination

In CPT, the Boundary primitive manages distinctions between self and others. Topics such as body schema, agency, and self-recognition (Dary & Lopez, 2023) are variations and elaborations of the Boundary primitive. Incorporating the primitive into the definition obviates the need to determine whether a species has self-recognition in the same sense as humans. Instead, it shifts the focus to how Boundary is realized across species. For example, membrane-level self/non-self discrimination in amoebae, self-generated versus externally caused stimulus discrimination in fish and insects, and elaborated body schemas and social self-representations in mammals can all be considered expressions of the same Boundary primitive. Since adaptive systems need the Boundary primitive (no adaptive system can persist without the ability to distinguish itself from its environment), questions about how species distinguish self from non-self become easier to study and answer.

While CPT may seem, at first glance, to do no more than describe complex behaviors in generalized terms by arranging every behavior and operation into one of six broad categories, a closer inspection shows that this is actually its strength. The breadth of the six primitives is what defines these operations at the computational level. In other words, the scale and generality of the primitives enable questions to be asked at the computational level rather than at the behavioral or mechanistic levels, making them easier and more accurate to categorize and compare across species. The five examples given illustrate this. CPT shifts the cross-species comparison question from whether different species exhibit certain elaborate behaviors or have specific neural or physical machinery to how the underlying primitives pair to operate and thus elaborate in each species. The argument here is not that measuring elaborate behaviors is wrong or unnecessary, or that they should not be studied across species. The point is that, at the level of cross-species comparison, primitive-level concepts translate more directly into a rigorous exploration of these comparative questions.

## 4. IMPLICATIONS OF THE SIX PRIMITIVES FOR ARTIFICIAL SYSTEMS

To this point, examples of the primitives have focused on biological organisms. However, the central claim of this theory is that the primitives are foundational across all adaptive systems and are therefore not necessarily limited to biological ones. To extend the concept of the primitives beyond biology, it is important to first justify this theoretical step.

Based on the information above, it is reasonable to infer that the computations appear in organisms that evolved independently. The computations therefore reflect general necessities of adaptation rather than features of any specific biological system. While this conclusion may be the best explanation given the evidence, it cannot be strictly argued from the premise. In other words, the theory and biological evidence support the claim that the computations are required by adaptation in general; they do not, on their own, establish that non-biological systems must also perform them. Such a claim requires further reasoning.

The bridge from biological to artificial systems must be built with care because, for one thing, the necessity criterion is grounded in survival. To qualify as an adaptive system, this criterion must be met. Machines, as ordinarily built, do not live, die, or reproduce. What generalizes to artificial systems, then, is not survival in the biological sense, but the more abstract concept of persistence, which gives rise to survival. For machines to persist, they must constantly maintain the conditions necessary for operational viability, meaning they must continue to function adaptively under environmental pressure. Additionally, they must be capable of failing when those conditions are not met. The necessity (*a*) criterion, in the substrate-neutral form defined above, is grounded in this possibility of failure. A machine, therefore, falls within the scope of this theory to the extent that it has something at stake. It must be in a condition where its continued functioning depends on its ability to actively maintain its state and can fail to do so. If a machine has nothing at stake, the primitives can, at best, be simulated. But where a machine must adapt to remain operationally viable, the primitives become necessary in the same sense that they are necessary for a biological organism to maintain itself.

Many artificial systems regulate variables, execute predefined responses, and can fail at such tasks without meeting the definition of persistence. A thermostat, for example, can maintain a set point and can also fail to do so. But its continued operation does not depend on its ability to hold the set point. That is because the thermostat has nothing at stake. The set point can be changed at any time in either direction without hindering the thermostat's ability to function as a thermostat. So, viability is not merely stipulated; it is determined by whether the conditions of its continued functioning are constitutive, meaning the system's operation depends on its ability to maintain them. Only when they are constitutive is anything genuinely at stake.

The difference between constitutive and stipulated conditions can be understood in biological autonomy, in which an organism's homeostatic norms are produced internally rather than imposed externally (Moreno & Mossio, 2015). The question is therefore not about a machine's ability to process information, but whether it can maintain the conditions of operational viability. CPT does not presume that current artificial systems meet the definition of persistence. Instead, it specifies what must be true of machines for the primitives to be applicable and suggests what the absence of the primitives would mean for systems built without them. Indeed, from a CPT perspective, many failures of current artificial systems reflect a lack of the computational primitives. Given this, two questions follow: Under what conditions can a machine be adaptive? And what does this theory imply about the machine intelligence currently being built?

To answer either question, the primitives must apply directly to machine intelligence. CPT presumes this and supplies what current theoretical approaches lack, namely, the computations required for machines to be adaptive systems. Machines are currently being built to perform specific tasks rather than to be adaptive (Goyal & Bengio, 2022), so they lack the very computations that adaptive existence requires. Thus, focusing broadly on adaptation rather than the narrow requirements of any particular task will realign the purposes and goals of artificial intelligence to more closely match the intelligence of biological systems. The advantage of such a realignment is not to replicate a biological substrate but to align with its shared computational processes.

Viewed this way, the primitives in artificial systems serve as a diagnostic tool rather than a direct explanation of cause. While the most common failures in current systems can each be explained by established engineering principles, the argument here is that, when considered collectively, these failures display the pattern expected of systems that lack the computations necessary for adaptive existence. Unlike approaches that propose separate solutions for each problem, this perspective holds that the primitives must work together as an integrated set. Implementing them individually, in a modular fashion as engineered solutions are designed to do, will not be sufficient. An adaptive system is greater than the sum of its computations, and those computations must function as a whole.

## 4.1 THE SUBSTRATE-NEUTRAL PRIMITIVES AND MACHINE INTELLIGENCE

Efforts to build more capable machine intelligence often draw on biology for inspiration (Hassabis et al., 2017; Zador et al., 2023). Biological systems are robust, general, and adaptive, which is the kind of intelligence engineers are trying to build into machines. An important question, then, is: What is it about biology that makes it a good model for building machine intelligence? How this question is answered determines what kind of system is built.

Perhaps the most common answer to this question is that biology's advantages lie in its machinery. The intuition is that a system modeled closely enough to the human brain will produce intelligence as a consequence. A main reason for modeling the human brain when building machine intelligence is to capture the flexible, transferable, general intelligence of biological systems that machines currently lack (Zador et al., 2023). Several programs pursue this intuition, from neuromorphic hardware (Schuman et al., 2022) to architectures modeled on cortical structure (Hawkins et al., 2019). These inquiries treat the brain's implementation of intelligence as the target to be reproduced. But trying to reproduce its neural machinery presumes that a brain-like structure is itself the source of intelligence. The problem with this approach is that the human brain's machinery is the product of a specific evolutionary history. It was shaped over hundreds of millions of years by the evolutionary demands of surviving and reproducing, all within the constraints of biochemistry and the varying environmental conditions it evolved in. Much of what the brain does is therefore incidental to the computations it performs. In other words, the human brain represents one solution for general intelligence that evolved in response to the demands placed on humans in their varying environments.

By contrast, CPT argues that the lesson from biology lies in the computational functions, not in the machinery that evolved to perform them. Evolution produced myriad ways to implement the computational primitives. The human brain is merely one example of an evolutionary solution for implementing them. The amoeba is another, as are the jellyfish, the bee, and the rat. If evolution works with what it has to perform these computations, whether that be a nervous system or a single cell, then a machine need not resemble the human brain to be an adaptive system any more than it needs to be built of biological-grade neurons. The only requirement is that it perform the computations that adaptive existence demands, even if a silicon-based system ultimately performs them.

Adopting this perspective reframes the question from "How do we make a system that is brain-like?" to "How do we make a system that operates according to the computational demands of adaptive existence?" The primitives enumerate those demands and provide the computations that meet them, thereby defining what a system must do to be adaptive. Thinking about machine intelligence this way places biology's contribution at the appropriate level: the computational level. At this level, the problems posed by adaptive existence can be stated and analyzed directly. The focus can then shift to understanding how a machine might solve them if it were required to adapt. This takes from biology what is generally and computationally related to intelligence and leaves behind the incidentals, such as the brain's specific functional machinery and its elaborate behaviors.

## 4.2 ADAPTIVE PROBLEMS MACHINE INTELLIGENCE MUST SOLVE

If adaptive existence indeed depends on these computational processes, any system lacking them should fail in predictable ways. Following, each primitive is examined in turn to see what failures it might predict. The adaptive problem it solves is explained, along with how current machine intelligence attempts to address it and how these issues relate to known types of failures in the field[7] (see Table 2).

[7] There are two important considerations when reading this subsection. First, the argument given is put forward to make predictions for interpretive purposes, not to give definitive evidence. Although current systems sometimes fail in ways consistent with a lack of the primitives, there are also established engineering explanations for these failures, so simply matching failure patterns to primitive descriptions is not sufficient to prove the point. Second, saying that each failure can be explained by a single

4.2.1 Arouse

The goal of Arouse is to align a system's effort with the demand it faces. Adaptive systems accomplish this by regulating how they allocate resources to tasks or situations. When demand is low, the system uses few resources. As demand increases, the system commits more and more resources to the task. When a system cannot regulate itself to these varying demands, it either exhausts itself by staying in a high-readiness state or remains undercommitted in a low-energy state, which may cause it to miss important opportunities or overlook threats.

**Table 2. The computational primitives in machine intelligence.** Each primitive is shown with the adaptive problem it solves, how it can be viewed in machine intelligence, and a documented failure in contemporary systems that the framework attributes to the absence of that primitive. The Computational Primitives Theory (CPT) holds that these failures, often treated as separate engineering problems with separate solutions, are the pattern one would expect when systems are built to perform specific tasks rather than to adapt, and therefore lack the computations that adaptive existence requires.

| Primitive | Adaptive Problem | In Machine Intelligence | Current Failure |
|---|---|---|---|
| ***Arouse*** | Adjust processing effort according to the demands of the situation | Distribute computational resources based on input requirements | Applies the same computational effort regardless of task difficulty; adaptive mechanisms are added after initial design (e.g., adaptive compute, inference-time scaling), not built-in |
| ***Orient*** | Identify and focus on inputs relevant to current goals while filtering out distractions | Prioritize information based on the system's objectives and current context | Prone to distraction; can be sidetracked by irrelevant or injected information; loses track over extended inputs |
| ***Valence*** | Assess what benefits or harms the system | Assign significance to stimuli in terms of approach or avoidance, grounded in the system's ongoing viability | Lacks intrinsic value; relies on external definitions, leading to reward hacking and specification gaming |
| ***Position*** | Determine the position of self and objects within a reference framework | Sustain a consistent understanding of self and the surrounding situation | Exhibits weak state tracking and inconsistent models of the world; agentic systems may lose their place during multi-step tasks |
| ***Boundary*** | Differentiate between self and non-self, and between familiar and unfamiliar information | Distinguish between grounded (factual) and generated content, as well as between in- and out-of-distribution data | Susceptible to hallucination and confabulation; vulnerable to prompt injection; lacks clear boundaries for reliable knowledge |
| ***Attune*** | Adapt to changing conditions as they are encountered over time | Fine-tune parameters in response to real-world conditions | Remains static after training; vulnerable to catastrophic forgetting; limited and short-lived in-context learning |

Current machine intelligence has recently begun to approach this kind of regulation, but it mainly focuses on engineered inference-time or routing mechanisms rather than native adaptive state variables primitive is a simplification deliberately chosen for explanatory purposes. In practice, a single failure almost assuredly results from more than one missing primitive. Therefore, the relationship between failures and primitives should be understood as one-to-many or many-to-many, not strictly one-to-one.

(Elhoushi et al., 2024; Yang et al., 2025; Bae et al., 2025). For example, a large language model (LLM) expends about the same compute on simple queries as on complex reasoning tasks (Schuster et al., 2022; Raposo et al., 2024). In transformer-based LLMs, the depth and width are fixed by the architecture, so the number of layers a token passes through remains the same regardless of how complex the task is. The cost of a single forward pass is the same regardless of whether the input is "what is two plus two" or involves multiple inferential steps.

The number of recent papers in confidence-based early exit strategies, mixture-of-depths routing, and test-time compute scaling indicates that the field is becoming increasingly aware of this issue and is looking for ways to balance the use of resources more selectively and judiciously (Schuster et al., 2022; Raposo et al., 2024; Snell et al., 2025). Nevertheless, problems related to the Arouse computation remain because these solutions are added post hoc to systems that already lack a native mechanism for matching resource allocation with task complexity.

#### 4.2.2 Orient

The purpose of Orient is to select and prioritize what is most important for downstream tasks. As a system receives input, it must filter and weight it according to its goals and current state. The input itself contains raw statistics that can be arranged hierarchically, but adaptive weighting cannot rely solely on the statistical structure of the inputs. Prioritization must align with what the system is trying to do and its current state.

When looking for the Orient computation in machines, the attention mechanism in modern neural networks (Vaswani et al., 2017) seems to compute Orient. It assigns different weights to inputs for each output. But that resemblance is misleading, because inputs influence attentional distributions only through fixed, learned transformations. Current systems lack a way to independently maintain a system's goal or viability state and adjust priorities as conditions change. Query and key projections learn their weights during training, and those weights are then fixed when the training ends. Thus, all future weighting depends on the surface input rather than on what the system is currently trying to do or the state it is in.

LLMs can represent goals, which seems like a viability-focused task, but those goals are fixed across inputs, and they lack an enduring system state with priority-setting authority. In other words, weighting depends on the input or feedback, not on what relates best to the system's viability state, which is what Orient is meant to compute. As a result, these systems are easily distracted by irrelevant context (Shi et al., 2023), injected instructions and prompts (Greshake et al., 2023), or by losing relevant information in long inputs (Liu et al., 2024). The problem, therefore, is not that current machines cannot weight inputs; it is that their weighting is not adaptive. Orient requires goal- and state-dependent prioritizations that fixed, statistically driven systems cannot provide.

#### 4.2.3 Valence

Valence provides the system with a valuation. As inputs arrive, the system computes whether those signals are good or bad based on these valuations. Those same valuations influence all subsequent outputs as well. In adaptive systems, both input and output valuations are grounded in the system's operational viability. Yet, while the Valence primitive is fundamental to adaptation, most current machine models lack its intrinsic evaluative capabilities and instead focus on optimizing externally provided valuations. Design engineers, reward models, preference data, and task objectives supply these valuations rather than having the system determine them from its own internal conditions of persistence (Christiano et al., 2017; Sutton & Barto, 2018; Ouyang et al., 2022).

According to CPT, Valence requires that valuation be grounded in the system's persistence. It is important to make this distinction so that alignment is not confused with Valence. Alignment tests how well the system conforms to human goals or preferences, so its valuation is tied to an external source. Valence, by contrast, requires the system to tie its valuations to its own persistence. Thus, alignment and Valence are related but not identical. A system can be well aligned while still lacking Valence, meaning it can conform to human valuations and goals without its own evaluative process. Research in adaptive robotics has addressed this directly. Architectures that derive behavior through internal viability rather than external rewards demonstrate that an agent's actions can be driven by its internal survival-related state (Lewis & Cañamero, 2016). Similarly, models that derive intrinsic motivation from information theory assess value based on the agent's ability to influence its own perceptions (Salge et al., 2014). An encouraging takeaway is that these methods suggest that endogenous evaluation can be built. Moreover, they suggest that evaluative behaviors of this kind must be intentionally designed, and that large-scale systems are not currently developed in this manner.

From another perspective, it has been argued that valuation based solely on reward maximization is sufficient to produce intelligence and its adaptive capacities (Silver et al., 2021). That argument, however, presupposes a reward supplied from outside the system. Rather than giving the system its own viability, it provides an external valuation that serves as a powerful optimizer, grounding the system in what is provided extrinsically. The system has no Valence of its own. This may be why specification gaming (Krakovna et al., 2020), along with reward hacking and misaligned optimization (Pan et al., 2022; Skalse et al., 2022), are so prevalent. In these cases, the machine optimizes a proxy—an evaluation supplied by an external reward model or objective function—for its own Valence rather than one that arises endogenously.

#### 4.2.4 Position

Position locates a system within the environment in which it operates and maintains a stable reference frame in that context. An adaptive system continuously computes Position to represent where it is in space at that moment, where other things are located, and how it actively relates to them, passing this information along to the other primitives for their computational use as well. For example, while memory (via Attune) stores and retrieves information, Position tracks where the system is in relation to that information, the task, and the environment. While Boundary distinguishes what belongs to the system, Position places the system within that structured field of relations.

In current machine intelligence, Position, as defined here, appears to be absent because most models fail to maintain a stable reference frame. As a result, models often lose track of where they are in a task, what state they and the world are in, and how their actions and outputs relate to the surrounding structure. This is evident in poor tracking performance over long interactions, where models fail to maintain consistent context across a conversation (Laban et al., 2025), and in the brittleness of agentic systems that lose track of where they are in a multi-step process (Yao et al., 2024). This can be traced to an architecture not built to maintain an intrinsic state or to keep its reference frame consistent across calls. Therefore, instead of retaining state, each step reconstructs the system's state from the tokens in the window and cached activation logs, or from external memory. Moreover, some of the leading agent architectures provide this continuity through interleaved reasoning-and-acting traces (Yao et al., 2023), external verbal memory and self-reflection (Shinn et al., 2023), or engineered memory streams with periodic reflection (Park et al., 2023). These findings further indicate that models construct and reconstruct a persistent frame of reference as they go. What is missing in all these failures is a persistent reference frame, meaning the system could not locate itself (what the Position primitive provides) relative to that frame. This explanation provides a better theoretical grounding than suggestions of any single isolated defect in memory or context handling.

#### 4.2.5 Boundary

The Boundary computation distinguishes what belongs to the system from what does not and identifies which information the system can process versus what falls outside its capabilities. The lack of this computation in machine intelligence is evident in failures of hallucination and confabulation (Huang et al., 2025). By definition, these failures point to the system's inability to understand what it does and does not know. It cannot differentiate between content grounded in fact and content that is not, and it has no reliable way of knowing what is in- or out-of-distribution (Yang et al., 2024).

Prompt injection is another example of Boundary failure. In this case, the system cannot distinguish its own instructions from those introduced externally (Greshake et al., 2023). Such a failure represents a form of cognitive predation,[8] in which an external agent captures and manipulates the system's processing precisely because the system cannot identify the information's source. It has no Boundary computation to solve this problem. While hallucination and prompt injection may seem distinct, with one concerned with distinguishing the known from the unknown (epistemic) and the other with recognizing its own information from external inputs (informational), they are both manifestations of the same underlying computational challenge that Boundary performs. Both ultimately reflect a failure to maintain a clear self/non-self distinction. In each case, what is missing is the capacity to maintain the boundaries the system needs to function, i.e., self vs. external, known vs. unknown, and in-domain vs. out-of-domain.

#### 4.2.6 Attune

Attune addresses the challenge of adapting to changing conditions over time. Adaptive systems continuously adjust their operational parameters based on what they encounter. Machine models, frozen after training, cannot. Adapting to new conditions requires extensive retraining, and even then, models often suffer from catastrophic forgetting, in which new learning overwrites old learning. Elastic weight consolidation and its successors are designed to limit this problem (Kirkpatrick et al., 2017). The concept is not new. Adaptive resonance theory frames it as the stability-plasticity dilemma: a system plastic enough to learn new information risks overwriting existing knowledge, while a stable system resists updating with new input (Grossberg, 2013). Attune is the computation that balances this trade-off in organisms, allowing them to adapt to new conditions without losing prior learning. In-context learning offers another solution, but it is a thin substitute that vanishes when the context window clears, leaving the underlying system unchanged (Brown et al., 2020). The deployed model is the one the user gets, regardless of what it subsequently encounters. The ability to adjust to new and changing conditions is missing. Requiring training to provide what may eventually be needed in deployment is one of the best arguments for an adaptive framework in machine intelligence.

The failures and shortcomings listed for each primitive are well-known limitations in machine learning, and each has its own proposed remedies. Problems around the uniformity of effort (Arouse) are addressed with adaptive computation; distractibility (Orient) with robustness research; misvalued objectives (Valence) with alignment research; loss of situational frame (Position) with memory and state-tracking work; hallucinations (Boundary) with retrieval and calibration methods; and frozen training models (Attune) with continual-learning research. Six different types of problems are being addressed with different solutions. The framework proposed here asserts that these problems are not unrelated but are instead six problems that arise when systems cannot adapt. Since these systems were not designed to be adaptive, they predictably lack the computations that adaptive existence requires. They were built to perform particularly well at certain tasks, and they do perform well at those tasks, but when they are placed in conditions that their training did not anticipate, and thus they did not train for, the absence of the primitives shows in the ways described above. These systems generalize poorly across domains (for example, Geirhos et al., 2020; Zhou et al., 2023), which is not surprising since cross-domain adaptation is precisely the capacity that the computational primitives provide.

[8] The term *cognitive predation* was suggested by R. Spencer Schaefer (personal communication, August 7, 2026). It is meant to describe an outside agent taking over a system's processing for its own purposes, which is possible in the machine context because the system cannot distinguish between self and other. The term is particularly apropos since predation and parasitism are failures of Boundary. For example, immune systems use self/non-self discrimination to defend against parasites. In artificial systems, a similar problem occurs, but without the natural defenses found in biology.

### 4.3 THE UNIFIED SET

Beyond the failures traced to each of the primitives above, this framework makes a broader, more consequential claim about the primitives: they must work together as a unified system. This claim puts forward a prediction that largely distinguishes CPT from the current field of artificial intelligence, where different problems are addressed in isolation. For example, solutions for adaptive computation, robustness, alignment, state-tracking, calibration, and continual learning are generally engineered independently, based on the tacit assumption that these capabilities can be added together at the end. The idea is that intelligence can be achieved by designing each skill separately and then integrating them into a single system.

CPT predicts the opposite. The primitives' contributions are not simply additive. A system cannot become adaptive by accumulating six computational competences engineered in isolation, because each primitive functions in relation to the others. Orient without Valence, for example, is not merely a degraded Orient; it is no longer Orient at all, since it lacks the full computational requirements of Orient. Thus, the prediction is that the primitives are not independently tunable. Improving one in isolation should not increase adaptive performance without the others. And if one is removed, it should degrade the adaptive performance of the remaining five, not merely subtract the contribution of the one removed. This prediction can be tested because the modular and unified perspectives differ on whether integrating separately engineered competencies leads to adaptation or merely reproduces the fragmentation observed in current models (Figure 2).

Testing this prediction requires a method to determine whether a machine computes a primitive rather than merely mimicking one, raising questions about empirical evidence for primitives in artificial systems. If the primitives are substrate-neutral computations rather than specific mechanisms, would it be obvious when a machine performed a primitive computation? How would one know? In other words, a system could be engineered to exhibit behavior that resembles a primitive computation while computing nothing of the kind. For example, a model might be engineered to perform a specific task that, on the surface, seems to attend to or evaluate something without performing anything close to the Orient or Valence computations. The challenge, then, is to clearly define what would count as evidence that a primitive has been computed. A practical test for machines is needed, similar to the one that guided the biological primitive candidate search. In that case, it was asked whether an amoeba could perform a certain function to judge whether the behavior was primary. What would the machine version of the "could an amoeba do it" test be?

One approach may be to evaluate the distinction between internally and externally imposed computation. In this test, a system would be judged to compute a primitive if its operation was based on its internal state rather than being provided during training, if it could generalize beyond its training data, and if it relied on its own conditions of persistence rather than an externally provided target. Any input-output response that was fixed and merely replicated a primitive's behavior on a benchmark test would not meet these criteria and therefore could not be considered computationally independent. While such a test would not be definitive, a result that meets these criteria would support a more robust interpretation than mere imitation. This is only a first approximation for validating the primitives in machines, but it does allow the question of how a machine computes the primitives to be asked in terms that are similar to those in biological systems.

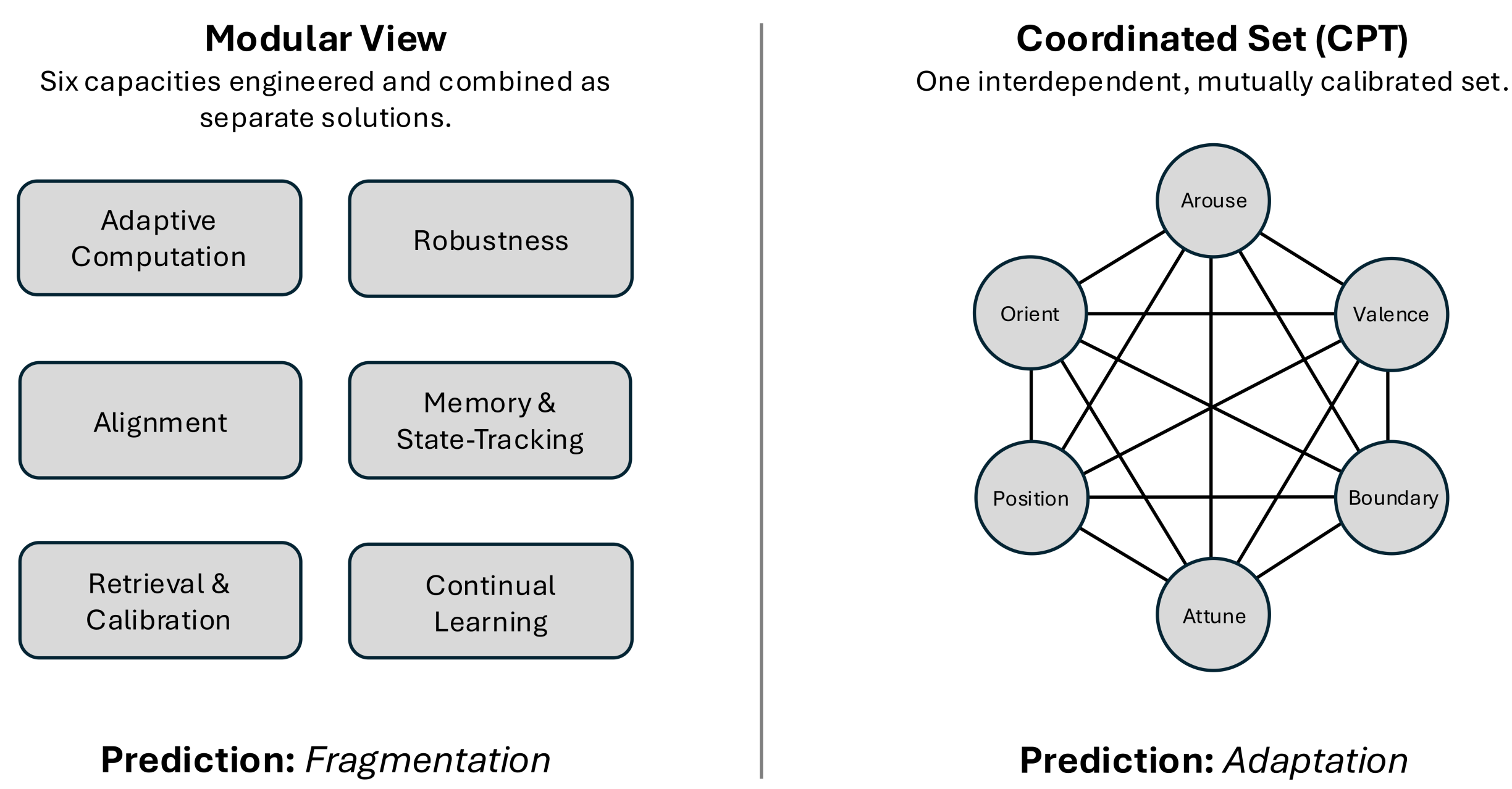

**Figure 2. The coordination prediction.** Two approaches to building adaptive systems are shown, each making a different prediction. The modular perspective (on the left) suggests that abilities such as robustness, alignment, state-tracking, calibration, and continual learning can be developed separately and then combined. By contrast, the Computational Primitives Theory (CPT) perspective (on the right) posits that these six primitives must operate collectively as an interconnected system. Because the two views predict different observable outcomes, either continued fragmentation or a cohesive adaptive system, the disagreement is empirically decidable.

Presenting tests like this poses potential challenges to the theory, not only through reinterpretation but also by suggesting that simple observations could contradict it. Such challenges can arise in two ways. The first challenge concerns the set's completeness. The theory's robustness would be compromised by any of the following: a system that is adaptive according to the given criteria yet lacks one of the six primitives; a seventh function that meets all four criteria but cannot be decomposed into the six primitives; or two proposed primitives that always co-occur and vary together, suggesting they are actually a single primitive (see Table 3). The second challenge is potentially the most crucial because it defines the framework's core commitment. The theory is false if the primitives can be added separately, in any order, with each functioning adaptively without the others. Stated another way, the theory would be false if incremental assembly, where each primitive performs its computations independently, led to an adaptive system. Such an outcome would align perfectly with the modular, task-focused approach to machine intelligence, making the comparison between these two approaches a straightforward test of the theory rather than just an interpretive issue.

**Table 3. Predicted selective dissociation of the primitives.** If the primitives are discrete computations, each should be able to fail without requiring the loss of the others, and each loss should produce a characteristic failure signature. This table outlines the expected consequences of losing each primitive separately in biological and artificial systems, with examples for artificial systems that mirror those in Table 2. Distinct, separable failure signatures indicate that the primitives are discrete computations. If any two are always lost or preserved together, it would suggest they are not separate computations.

| Primitive | Selective loss in biological systems | Selective loss/absence in artificial systems |
|---|---|---|
| ***Arouse*** | Failure to mobilize or down-regulate energy and activity to match demand (pathological fatigue; hypo- or hyper-arousal) | Uniform computation regardless of difficulty, no adaptive allocation of compute (same effort on trivial and hard inputs) |
| ***Orient*** | Failure to select relevant inputs; neglect or pathological distractibility | Inability to direct processing to relevant information; degradation and distractibility over long contexts |
| ***Valence*** | Loss of approach/avoid evaluation; inability to distinguish beneficial from harmful (indiscriminate approach; anhedonia) | Reward hacking and specification gaming, optimizing a proxy with no grounded evaluation of its own |
| ***Position*** | Spatial and self-locational disorientation; failure to localize one's self relative to environment (topographical disorientation) | Loss of state/context tracking, failure to maintain "where am I" in the task or environment (context drift) |
| ***Boundary*** | Breakdown of self/non-self discrimination, autoimmunity, or failure to distinguish self-generated from external signals | Hallucination and confabulation (known vs. unknown) and prompt injection (own vs. external instructions) |
| ***Attune*** | Failure to update responses from experience, or failure to retain what is learned | Catastrophic forgetting, new learning overwrites prior competence |

Taken together, these failures and the prediction they frame open a new line of thinking for building machine intelligence. They shift the goal from producing systems that perform specific tasks to producing systems that compute what adaptive existence requires. The framework does not yet answer these questions empirically, but it specifies what the questions are, why they should be asked, and, in the unified set prediction, at least one outcome that would show it to be wrong.

## 5. CONCLUSION

This paper began with a question (Q3) that is often overlooked in the study of adaptive systems. Beyond what an organism does and how its machinery works, there is the question of what it must compute to persist. Taking that third question as the starting point, the theory proposed here holds that adaptive existence rests on a small, specifiable set of computational primitives: Arouse, Orient, Valence, Position, Boundary, and Attune. Each is necessary for survival, recurs across independently evolved lineages, is conserved in the machinery of living organisms, and is irreducible to the others. Defined at the level of computation rather than implementation, these primitives are offered as answers to what any system must compute to survive and propagate. The honeybee, whose flight and dance were described at the beginning of this paper, can now be described not only by what she does and how her neural machinery accomplishes it, but also by what she is computing to persist.

Defining the primitives this way means they are not tied to biological systems. The four criteria were designed to select functions that hold regardless of the substrate that performs them, so the same set of primitives that describes an amoeba, a bee, or a human should also describe any system that must adapt to persist, including adaptive machines. For biological systems, this offers a level of analysis beneath behavior and mechanism, allowing distantly related organisms to be compared without forcing one species' concepts onto another. For artificial systems, it suggests that common failures in current machine intelligence may not be merely engineering flaws but instead reflect a behavioral pattern expected in machines built to perform certain tasks rather than adapt. In this way, CPT serves the dual purpose of comparison and construction.

Thus, a single set of computational primitives underpins both biological and artificial systems. That is CPT's central claim. And since carbon- and silicon-based systems occupy the same physical and spatial world and therefore could face the same demands for persistence, it is posited that the computational primitives are the operations that both require. Given this, the difference between an organism and a machine shifts from what each system requires to be adaptive to determining whether either or both substrates supply it. If a machine does compute the primitives as a coordinated set, it would not necessarily be imitating life; rather, it would be doing what living systems do, using the same computational logic, just with a different physical material.

When considering the claims made in this paper collectively, its main contribution may be seen as being more about adaptive systems than about the primitives. Thus, it may be more precise to view the primitives as computations of adaptation rather than as an exhaustive list. They may therefore better serve as an initial classification of operations, with their most important contribution being establishing this level of adaptation. If the main idea, as in Marr's levels of analysis, is that adaptation can be characterized in computational terms regardless of the substrate, then the primitives may serve better as specific, testable proposals that make this level of adaptation feasible. If that is the case, then recognizing adaptation as a distinct computational domain is what matters, even if the particular decomposition of the primitives described here is later revised, since it is the level of adaptation, not the particular list of primitives, that shapes the relevant questions around it.

What is offered, then, is a working hypothesis. The set of six computational primitives may prove incomplete; the discriminatory borders between them may need to be redrawn; and the four criteria themselves may face scrutiny. These are all good things. They are the appropriate kinds of issues to raise because they are answerable through argument, empirical testing, comparative evidence, and, ultimately, attempts to implement the primitives in machines that lack them. The purpose of this paper is to specify what the computations are and why they belong together, and, in doing so, to suggest that adaptation, wherever it occurs, runs on a common set of operations that are capable of making machines adaptive too, once they have been built to compute them.

ACKOWLEDGEMENTS

I thank R. Spencer Schaefer, PharmD, for his critical review of the manuscript and for several extensive discussions about the arguments presented in the artificial intelligence sections, particularly those concerning Boundary and what it means for a machine to fail. His suggestions substantially improved the manuscript.

I also thank Orrin Page, PhD, for helping me work through the logical arguments presented here, for tightening the definitions and descriptions of the primitives on paper and in my mind, and for editing the manuscript. His careful attention to the structure of the argument and the text improved both.

REFERENCES

Alessandroni, N., Altschul, D., Bazhydai, M., Byers-Heinlein, K., Elsherif, M., Gjoneska, B., Huber, L., Mazza, V., Miller, R., Nawroth, C., Pronizius, E., Qadri, M. A. J., Šlipogor, V., Soderstrom, M., Stevens, J. R., Visser, I., Williams, M., Zettersten, M., & Prétôt, L. (2024). Comparative cognition needs Big Team Science: How large-scale collaborations will unlock the future of the field. *Comparative Cognition & Behavior Reviews, 19*, 67–72. https://doi.org/10.3819/CCBR.2024.190001

Anderson, D. J., & Adolphs, R. (2014). A framework for studying emotions across species. *Cell, 157*(1), 187–200. https://doi.org/10.1016/j.cell.2014.03.003

Aso, Y., Sitaraman, D., Ichinose, T., Kaun, K. R., Vogt, K., Belliart-Guérin, G., Plaçais, P.-Y., Robie, A. A., Yamagata, N., Schnaitmann, C., Rowell, W. J., Johnston, R. M., Ngo, T.-T. B., Chen, N., Korff, W., Nitabach, M. N., Heberlein, U., Preat, T., Branson, K. M., Tanimoto, H., & Rubin, G. M. (2014). Mushroom body output neurons encode valence and guide memory-based action selection in *Drosophila*. *eLife, 3,* e04580. https://doi.org/10.7554/eLife.04580

Aston-Jones, G., & Cohen, J. D. (2005). An integrative theory of locus coeruleus-norepinephrine function: Adaptive gain and optimal performance. *Annual Review of Neuroscience, 28,* 403–450. https://doi.org/10.1146/annurev.neuro.28.061604.135709

Bae, S., Kim, Y., Bayat, R., Kim, S., Ha, J., Schuster, T., Fisch, A., Harutyunyan, H., Ji, Z., Courville, A., & Yun, S.-Y. (2025). *Mixture-of-Recursions: Learning Dynamic Recursive Depths for Adaptive Token-Level Computation*. arXiv. https://doi.org/10.48550/arXiv.2507.10524

Baier, H. & Scott, E. K. (2024). The visual systems of zebrafish. *Annual Review of Neuroscience*, *47,* 255–276. https://doi.org/10.1146/annurev-neuro-111020-104854

Baratti, G., Potrich, D., Lee, S. A., Morandi-Raikova, A., & Sovrano, V. A. (2022). The geometric world of fishes: A synthesis on spatial reorientation in teleosts. *Animals, 12*(7), 881. https://doi.org/10.3390/ani12070881

Bianco, I. H., & Engert, F. (2015). Visuomotor transformations underlying hunting behavior in zebrafish. *Current Biology, 25*(7), 831–846. https://doi.org/10.1016/j.cub.2015.01.042

Boisseau, R. P., Vogel, D., & Dussutour, A. (2016). Habituation in non-neural organisms: Evidence from slime moulds. *Proceedings of the Royal Society B: Biological Sciences, 283*(1829), 20160446. https://doi.org/10.1098/rspb.2016.0446

Boussard, A., Fessel, A., Oettmeier, C., Briard, L., Döbereiner, H.-G., & Dussutour, A. (2021). Adaptive behaviour and learning in slime moulds: The role of oscillations. *Philosophical Transactions of the Royal Society B: Biological Sciences, 376*(1820), 20190757. https://doi.org/10.1098/rstb.2019.0757

Bouton, M. E., Maren, S., & McNally, G. P. (2021). Behavioral and neurobiological mechanisms of Pavlovian and instrumental extinction learning. *Physiological Reviews, 101*(2), 611–681. https://doi.org/10.1152/physrev.00016.2020

Bräuer, J., Hanus, D., Pika, S., Gray, R., & Uomini, N. (2020). Old and new approaches to animal cognition: There is not "one cognition." *Journal of Intelligence, 8*(3), 28. https://doi.org/10.3390/jintelligence8030028

Brown, T. B., Mann, B., Ryder, N., Subbiah, M., Kaplan, J. D., Dhariwal, P., Neelakantan, A., Shyam, P., Sastry, G., Askell, A., Agarwal, S., Herbert-Voss, A., Krueger, G., Henighan, T., Child, R., Ramesh, A., Ziegler,

D. M., Wu, J., Winter, C., ... Amodei, D. (2020). Language models are few-shot learners. In *Advances in Neural Information Processing Systems 33* (pp. 1877–1901). Curran Associates.
Bshary, R., & Brown, C. (2014). Fish cognition. *Current Biology, 24*(19), R947–R950. https://doi.org/10.1016/j.cub.2014.08.043
Caporale, N., & Dan, Y. (2008). Spike timing-dependent plasticity: a Hebbian learning rule. *Annual Review of Neuroscience, 31*, 25-46. https://doi.org/10.1146/annurev.neuro.31.060407.125639
Castro, F., Lenggenhager, B., Zeller, D., Pellegrino, G., D'Alonzo, M., & Di Pino, G. (2023). From rubber hands to neuroprosthetics: Neural correlates of embodiment. *Neuroscience & Biobehavioral Reviews, 153*, 105351. https://doi.org/10.1016/j.neubiorev.2023.105351
Chen, W. (2022). Neural circuits provide insights into reward and aversion. *Frontiers in Neural Circuits, 16*, 1002485. https://doi.org/10.3389/fncir.2022.1002485
Christiano, P. F., Leike, J., Brown, T. B., Martic, M., Legg, S., & Amodei, D. (2017). Deep reinforcement learning from human preferences. In *Advances in Neural Information Processing Systems 30* (pp. 4299–4307). Curran Associates.
Colin, R., Ni, B., Laganenka, L. & Sourjik, V. (2021). Multiple functions of flagellar motility and chemotaxis in bacterial physiology. *FEMS Microbiology Reviews*, 45(6), fuab038. https://doi.org/10.1093/femsre/fuab038
Collett, M., Chittka, L., & Collett, T. S. (2013). Spatial memory in insect navigation. *Current Biology, 23*(17), R789–R800. https://doi.org/10.1016/j.cub.2013.07.020
Crapse, T. B., & Sommer, M. A. (2008). Corollary discharge across the animal kingdom. *Nature Reviews Neuroscience, 9*, 587–600. https://doi.org/10.1038/nrn2457
Daly, K. C., & Dacks, A. (2023). The self as part of the sensory ecology: How behavior affects sensation from the inside out. *Current Opinion in Insect Science, 58*, 101053. https://doi.org/10.1016/j.cois.2023.101053
Dary, Z., & Lopez, C. (2023). Understanding the neural bases of bodily self-consciousness: Recent achievements and main challenges. *Frontiers in Integrative Neuroscience, 17*, 1145924. https://doi.org/10.3389/fnint.2023.1145924
Davies, J. R., & Clayton, N. S. (2024). Is episodic-like memory *like* episodic memory? *Philosophical Transactions of the Royal Society B: Biological Sciences, 379*(1913), 20230397. https://doi.org/10.1098/rstb.2023.0397
de Bivort, B. L., & van Swinderen, B. (2016). Evidence for selective attention in the insect brain. *Current Opinion in Insect Science, 15,* 9–15. https://doi.org/10.1016/j.cois.2016.02.007
de Oliveira Mann, C. C., & Hornung, V. (2021). Molecular mechanisms of nonself nucleic acid recognition by the innate immune system. *European Journal of Immunology, 51*(8), 1897–1910. https://doi.org/10.1002/eji.202049116
Debanne, D., & Inglebert, Y. (2023). Spike timing-dependent plasticity and memory. *Current Opinion in Neurobiology, 80*, 102707. https://doi.org/10.1016/j.conb.2023.102707
Di Paolo, E. A. (2005). Autopoiesis, adaptivity, teleology, agency. *Phenomenology and the Cognitive Sciences, 4*(4), 429–452. https://doi.org/10.1007/s11097-005-9002-y
Domenici, P., & Hale, M. E. (2019). Escape responses of fish: A review of the diversity in motor control, kinematics and behaviour. *Journal of Experimental Biology, 222*(18), jeb166009. https://doi.org/10.1242/jeb.166009
Dong, L. L., & Fiete, I. R. (2024). Grid cells in cognition: Mechanisms and function. *Annual Review of Neuroscience, 47*, 345–368. https://doi.org/10.1146/annurev-neuro-101323-112047
Dussutour, A. (2021). Learning in single cell organisms. *Biochemical and Biophysical Research Communications, 564*, 92–102. https://doi.org/10.1016/j.bbrc.2021.02.018
El-Sobky, M. H., Rijal, R., & Gomer, R. H. (2025). Two endogenous *Dictyostelium discoideum* chemorepellents use different mechanisms to induce repulsion. *Proceedings of the National Academy of the Sciences, U.S.A., 122*(22) e2503168122. https://doi.org/10.1073/pnas.2503168122
Elhoushi, M., Shrivastava, A., Likhomanenko, T., Ozlem, Y., Haziza, L., Stojnic, N., ... & Dubey, A. (2024). LayerSkip: Enabling early exit inference and self-speculative decoding. *Proceedings of the 62nd Annual Meeting of the Association for Computational Linguistics (Volume 1: Long Papers)*, 5972–5991. https://doi.org/10.48550/arXiv.2404.16710
Evans, B. J. E., O'Carroll, D. C., Fabian, J. M. & Wiederman, S. D. (2022). Dragonfly neurons selectively attend to targets within natural scenes. *Frontiers in Cellular Neuroscience, 16,* 857071. https://doi.org/10.3389/fncel.2022.857071
Feldman, H., & Friston, K. J. (2010). Attention, uncertainty, and free-energy. *Frontiers in Human Neuroscience, 4*, 215. https://doi.org/10.3389/fnhum.2010.00215

Finke, V., Scheiner, R., Giurfa, M., & Avarguès-Weber, A. (2023). Individual consistency in the learning abilities of honey bees: Cognitive specialization within sensory and reinforcement modalities. *Animal Cognition, 26*(3), 909–928. https://doi.org/10.1007/s10071-022-01741-2

Freas, C. A., & Cheng, K., (2022). The basis of navigation across species. *Annual Review of Psychology, 73*, 217–241, https://doi.org/10.1146/annurev-psych-020821-111311

Freas, C. A., & Spetch, M. L., (2023). Varieties of visual navigation in insects. *Animal Cognition, 26*(1), 319–342, https://doi.org/10.1007/s10071-022-01720-7

Friston, K. (2010). The free-energy principle: A unified brain theory? *Nature Reviews Neuroscience, 11*(2), 127–138. https://doi.org/10.1038/nrn2787

Friston, K., FitzGerald, T., Rigoli, F., Schwartenbeck, P., & Pezzulo, G. (2017). Active inference: A process theory. *Neural Computation, 29*(1), 1-49. https://doi.org/10.1162/NECO_a_00912

Gallup, G. G., Jr. (1970). Chimpanzees: Self-recognition. *Science, 167*(3914), 86–87. https://doi.org/10.1126/science.167.3914.86

Gardner, R. A., & Gardner, B. T. (1969). Teaching sign language to a chimpanzee. *Science, 165*(3894), 664–672. https://doi.org/10.1126/science.165.3894.664

Geirhos, R., Jacobsen, J.-H., Michaelis, C., Zemel, R., Brendel, W., Bethge, M., & Wichmann, F. A. (2020). Shortcut learning in deep neural networks. *Nature Machine Intelligence, 2*(11), 665–673. https://doi.org/10.1038/s42256-020-00257-z

Geva-Sagiv, M., Las, L., Yovel, Y., & Ulanovsky, N. (2015). Spatial cognition in bats and rats: From sensory acquisition to multiscale maps and navigation. *Nature Reviews Neuroscience, 16*, 94–108. https://doi.org/10.1038/nrn3888

Ghose, D., Elston, T., & Lew, D. (2022). Orientation of cell polarity by chemical gradients. *Annual Review of Biophysics, 51*, 431–451. https://doi.org/10.1146/annurev-biophys-110821-071250

Gibson, W. T., Gonzalez, C. R., Fernandez, C., Ramasamy, L., Tabachnik, T., Du, R. R., Felsen, P. D., Maire, M. R., Perona, P., & Anderson, D. J. (2015). Behavioral responses to a repetitive visual threat stimulus express a persistent state of defensive arousal in *Drosophila*. *Current Biology, 25*(11), 1401–1415. https://doi.org/10.1016/j.cub.2015.03.058

Glanzman, D. L. (2010). Common mechanisms of synaptic plasticity in vertebrates and invertebrates. *Current Biology, 20*(1), R31-R36. https://doi.org/10.1016/j.cub.2009.10.023

Goyal, A., & Bengio, Y. (2022). Inductive biases for deep learning of higher-level cognition. *Proceedings of the Royal Society A: Mathematical, Physical and Engineering Sciences, 478*(2266), Article 20210068. https://doi.org/10.1098/rspa.2021.0068

Greshake, K., Abdelnabi, S., Mishra, S., Endres, C., Holz, T., & Fritz, M. (2023). Not what you've signed up for: Compromising real-world LLM-integrated applications with indirect prompt injection. In *Proceedings of the 16th ACM Workshop on Artificial Intelligence and Security (AISec '23)* (pp. 79–90). Association for Computing Machinery. https://doi.org/10.1145/3605764.3623985

Grima, L. L., Haberkern, H., Mohanta, R., Morimoto, M. M., Rajagopalan, A. E., & Scholey, E. V. (2025). Foraging as an ethological framework for neuroscience. *Trends in Neurosciences, 48*(11), 877–890. https://doi.org/10.1016/j.tins.2025.08.006

Gross, J. D. & Pears, C. J. (2021). Possible involvement of the nutrient and energy sensors mTORC1 and AMPK in cell fate diversification in a non-metazoan organism. *Frontiers in Cell and Developmental Biology*, 9, 758317. https://doi.org/10.3389/fcell.2021.758317

Grossberg, S. (2013). Adaptive resonance theory: How a brain learns to consciously attend, learn, and recognize a changing world. *Neural Networks, 37,* 1–47. https://doi.org/10.1016/j.neunet.2012.09.017

Guo, H. & Dixon, B. (2021). Understanding acute stress-mediated immunity in teleost fish. *Fish and Shellfish Immunology Reports*, 2, 100010. https://doi.org/10.1016/j.fsirep.2021.100010

Guo, L., Wang, Y.-H., Cui, R., Huang, Z., Hong, Y., Qian, J.-W., Ni, B., Xu, A.-M., Jiang, C.-Y., Zhulin, I. B., Liu, S.-J. & Li, D.-F. (2023). Attractant and repellent induce opposing changes in the four-helix bundle ligand-binding domain of a bacterial chemoreceptor. *PLOS Biology*, 21(12), e3002429. https://doi.org/10.1371/journal.pbio.3002429

Hafed, Z. M., Hoffmann, K.-P., Chen, C.-Y. & Bogadhi, A. R. (2023). Visual functions of the primate superior colliculus. *Annual Review of Vision Science, 9,* 361–383. https://doi.org/10.1146/annurev-vision-111022-123817

Hagihara, K. M., & Lüthi, A. (2024). Bidirectional valence coding in amygdala intercalated clusters: A neural substrate for the opponent-process theory of motivation. *Neuroscience Research, 209*, 28–33. https://doi.org/10.1016/j.neures.2024.07.003

Hao, M., Li, F., Duan, J.-W. & Han, M.-H. (2025). Neural circuit connections and functions of locus coeruleus–norepinephrine system. *International Journal of Molecular Sciences*, 26(22), 11163. https://doi.org/10.3390/ijms262211163

Hassabis, D., Kumaran, D., Summerfield, C., & Botvinick, M. (2017). Neuroscience-inspired artificial intelligence. *Neuron, 95*(2), 245–258. https://doi.org/10.1016/j.neuron.2017.06.011

Hawkins, J., Lewis, M., Klukas, M., Purdy, S., & Ahmad, S. (2019). A framework for intelligence and cortical function based on grid cells in the neocortex. *Frontiers in Neural Circuits, 12*, Article 121. https://doi.org/10.3389/fncir.2018.00121

Hazelbauer, G. L., Falke, J. J., & Parkinson, J. S. (2008). Bacterial chemoreceptors: High-performance signaling in networked arrays. *Trends in Biochemical Sciences, 33*(1), 9–19. https://doi.org/10.1016/j.tibs.2007.09.014

Hopper, L. M., Gulli, R. A., Howard, L. H., Kano, F., Krupenye, C., Ryan, A. M., & Paukner, A. (2021). The application of noninvasive, restraint-free eye-tracking methods for use with nonhuman primates. *Behavior Research Methods, 53*(3), 1003–1030. https://doi.org/10.3758/s13428-020-01465-6

Huang, L., Yu, W., Ma, W., Zhong, W., Feng, Z., Wang, H., Chen, Q., Peng, W., Feng, X., Qin, B., & Liu, T. (2025). A survey on hallucination in large language models: Principles, taxonomy, challenges, and open questions. *ACM Transactions on Information Systems, 43*(2), Article 42. https://doi.org/10.1145/3703155

Hulse, B. K., Haberkern, H., Franconville, R., Turner-Evans, D., Takemura, S.-Y., Wolff, T., Noorman, M., Dreher, M., Dan, C., Parekh, R., Hermundstad, A. M., Rubin, G. M., & Jayaraman, V. (2021). A connectome of the *Drosophila* central complex reveals network motifs suitable for flexible navigation and context-dependent action selection. *eLife, 10*, e66039. https://doi.org/10.7554/eLife.66039

Iglesias, P. A., & Devreotes, P. N. (2008). Navigating through models of chemotaxis. *Current Opinion in Cell Biology, 20*(1), 35–40. https://doi.org/10.1016/j.ceb.2007.11.011

Insall, R. H. (2010). Understanding eukaryotic chemotaxis: A pseudopod-centred view. *Nature Reviews Molecular Cell Biology, 11*(6), 453–458. https://doi.org/10.1038/nrm2905

Insall, R. H., Paschke, P. & Tweedy, L. (2022). Steering yourself by the bootstraps: How cells create their own gradients for chemotaxis. *Trends in Cell Biology*, 32(7), 585–596. https://doi.org/10.1016/j.tcb.2022.02.007

Jakowec, N. & Finkel, S. E. (2025). Controlled burn: interconnections between energy-spilling pathways and metabolic signaling in bacteria. *Journal of Bacteriology*, 207(4), e00542-24. https://doi.org/10.1128/jb.00542-24

Janak, P. H., & Tye, K. M. (2015). From circuits to behaviour in the amygdala. *Nature, 517*(7534), 284–292. https://doi.org/10.1038/nature14188

Janetopoulos, C., & Firtel, R. A. (2008). Directional sensing during chemotaxis. *FEBS Letters, 582*(14), 2075–2085. https://doi.org/10.1016/j.febslet.2008.04.035

Janeway, C. A., Jr., & Medzhitov, R. (2002). Innate immune recognition. *Annual Review of Immunology, 20,* 197–216. https://doi.org/10.1146/annurev.immunol.20.083001.084359

Jeffery, K. J., Cheng, K., Newcombe, N. S., Bingman, V. P., & Menzel, R. (2024). Unpacking the navigation toolbox: Insights from comparative cognition. *Proceedings of the Royal Society B: Biological Sciences, 291*(2016), 20231304. https://doi.org/10.1098/rspb.2023.1304

Joshi, S., Li, Y., Kalwani, R. M., & Gold, J. I. (2016). Relationships between pupil diameter and neuronal activity in the locus coeruleus, colliculi, and cingulate cortex. *Neuron, 89*(1), 221–234. https://doi.org/10.1016/j.neuron.2015.11.028

Kandel, E. R. (2001). The molecular biology of memory storage: A dialogue between genes and synapses. *Science, 294*(5544), 1030–1038. https://doi.org/10.1126/science.1067020

Kannan, K., Galizia, C. G. & Nouvian, M. (2022). Olfactory strategies in the defensive behaviour of insects. *Insects*, 13(5), 470. https://doi.org/10.3390/insects13050470

Kay, R. R. (2021). Macropinocytosis: Biology and mechanisms. *Cells & Development, 168*, 203713. https://doi.org/10.1016/j.cdev.2021.203713

Keizer-Gunnink, I., Kortholt, A., & Van Haastert, P. J. M. (2007). Chemoattractants and chemorepellents act by inducing opposite polarity in phospholipase C and PI3-kinase signaling. *The Journal of Cell Biology, 177*(4), 579–585. https://doi.org/10.1083/jcb.200611046

Kirchhoff, M., Parr, T., Palacios, E., Friston, K., & Kiverstein, J. (2018). The Markov blankets of life: Autonomy, active inference and the free energy principle. *Journal of the Royal Society Interface, 15*(138), 20170792. https://doi.org/10.1098/rsif.2017.0792

Kirkpatrick, J., Pascanu, R., Rabinowitz, N., Veness, J., Desjardins, G., Rusu, A. A., Milan, K., Quan, J., Ramalho, T., Grabska-Barwinska, A., Hassabis, D., Clopath, C., Kumaran, D., & Hadsell, R. (2017). Overcoming catastrophic forgetting in neural networks. *Proceedings of the National Academy of Sciences, 114*(13), 3521–3526. https://doi.org/10.1073/pnas.1611835114
Kirolos, S. A., & Gomer, R. H. (2022). A chemorepellent inhibits local Ras activation to inhibit pseudopod formation to bias cell movement away from the chemorepellent. *Molecular Biology of the Cell, 33*(1), ar9. https://doi.org/10.1091/mbc.e20-10-0656
Knudsen, E. I. (2007). Fundamental components of attention. *Annual Review of Neuroscience, 30,* 57–78. https://doi.org/10.1146/annurev.neuro.30.051606.094256
Kobayashi, T., Kohda, M., Awata, S., Bshary, R., & Sogawa, S. (2024). Cleaner fish with mirror self-recognition capacity precisely realize their body size based on their mental image. *Scientific Reports, 14*, 20202. https://doi.org/10.1038/s41598-024-70138-7
Kohda, M., Hotta, T., Takeyama, T., Awata, S., Tanaka, H., Asai, J.-y., & Jordan, A. L. (2019). If a fish can pass the mark test, what are the implications for consciousness and self-awareness testing in animals? *PLOS Biology, 17*(2), e3000021. https://doi.org/10.1371/journal.pbio.3000021
Kohda, M., Bshary, R., Kubo, N., & Sogawa, S. (2023). Cleaner fish recognize self in a mirror via self-face recognition like humans. *PNAS, 120*(7), e2208420120, https://doi.org/10.1073/pnas.2208420120
Krakovna, V., Uesato, J., Mikulik, V., Rahtz, M., Everitt, T., Kumar, R., Kenton, Z., Leike, J., & Legg, S. (2020, April 21). *Specification gaming: The flip side of AI ingenuity*. Google DeepMind. https://deepmind.google/blog/specification-gaming-the-flip-side-of-ai-ingenuity/
Krauzlis, R. J., Lovejoy, L. P., & Zénon, A. (2013). Superior colliculus and visual spatial attention. *Annual Review of Neuroscience, 36,* 165–182. https://doi.org/10.1146/annurev-neuro-062012-170249
Krauzlis, R. J., Wang, L., Yu, G., & Katz, L. N. (2023). What is attention? *WIREs Cognitive Science, 14*(1), e1570. https://doi.org/10.1002/wcs.1570
Kuhn, J., Lin, Y., & Devreotes, P. N. (2021). Using live-cell imaging and synthetic biology to probe directed migration in *Dictyostelium*. *Frontiers in Cell and Developmental Biology, 9*, 740205. https://doi.org/10.3389/fcell.2021.740205
Laban, P., Hayashi, H., Zhou, Y., & Neville, J. (2025). *LLMs get lost in multi-turn conversation*. arXiv. https://doi.org/10.48550/arXiv.2505.06120
Lei, Y. (2023). Sociality and self-awareness in animals. *Frontiers in Psychology, 13*, 1065638. https://doi.org/10.3389/fpsyg.2022.1065638
Levin, M. (2022). Technological approach to mind everywhere: An experimentally-grounded framework for understanding diverse bodies and minds. *Frontiers in Systems Neuroscience, 16*, 768201. https://doi.org/10.3389/fnsys.2022.768201
Lewis, M., & Cañamero, L. (2016). Hedonic quality or reward? A study of basic pleasure in homeostasis and decision making of a motivated autonomous robot. *Adaptive Behavior, 24(5),* 267–291. https://doi.org/10.1177/1059712316666331
Lindsay, G. W. (2020). Attention in psychology, neuroscience, and machine learning. *Frontiers in Computational Neuroscience, 14*, 29. https://doi.org/10.3389/fncom.2020.00029
Liu, N. F., Lin, K., Hewitt, J., Paranjape, A., Bevilacqua, M., Petroni, F., & Liang, P. (2024). Lost in the middle: How language models use long contexts. *Transactions of the Association for Computational Linguistics, 12*, 157–173. https://doi.org/10.1162/tacl_a_00638
Lyon, P. (2015). The cognitive cell: Bacterial behavior reconsidered. *Frontiers in Microbiology, 6,* 264. https://doi.org/10.3389/fmicb.2015.00264
Lyon, P., Keijzer, F., Arendt, D., & Levin, M. (2021). Reframing cognition: Getting down to biological basics. *Philosophical Transactions of the Royal Society B, 376*(1820), 20190750. https://doi.org/10.1098/rstb.2019.0750
Malenka, R. C., & Bear, M. F. (2004). LTP and LTD: An embarrassment of riches. *Neuron, 44*(1), 5–21. https://doi.org/10.1016/j.neuron.2004.09.012
Maniak, M. (2002). Conserved features of endocytosis in *Dictyostelium*. *International Review of Cytology, 221,* 257–287. https://doi.org/10.1016/S0074-7696(02)21014-1
Maravita, A., & Iriki, A. (2004). Tools for the body (schema). *Trends in Cognitive Sciences, 8*(2), 79–86. https://doi.org/10.1016/j.tics.2003.12.008
Marr, D. (1982). *Vision: A computational investigation into the human representation and processing of visual information*. W. H. Freeman.
Matsumoto, N., Barson, D., Liang, L., & Crair, M. C. (2024). Hebbian instruction of axonal connectivity by endogenous correlated spontaneous activity. *Science, 385*(6710), eadh7814. https://doi.org/10.1126/science.adh7814

Maturana, H. R., & Varela, F. J. (1980). *Autopoiesis and cognition: The realization of the living*. D. Reidel.
McNaughton, B. L., Battaglia, F. P., Jensen, O., Moser, E. I., & Moser, M.-B. (2006). Path integration and the neural basis of the 'cognitive map'. *Nature Reviews Neuroscience, 7*(8), 663–678. https://doi.org/10.1038/nrn1932
Menzel, R. (2012). The honeybee as a model for understanding the basis of cognition. *Nature Reviews Neuroscience, 13*(11), 758–768. https://doi.org/10.1038/nrn3357
Mobbs, D., Trimmer, P. C., Blumstein, D. T., & Dayan, P. (2018). Foraging for foundations in decision neuroscience: Insights from ethology. *Nature Reviews Neuroscience, 19*, 419-427. https://doi.org/10.1038/s41583-018-0010-7
Modi, M. N., Shuai, Y., & Turner, G. C. (2020). The *Drosophila* mushroom body: From architecture to algorithm in a learning circuit. *Annual Review of Neuroscience, 43*, 465–484. https://doi.org/10.1146/annurev-neuro-080317-0621333
Moreno, A., & Mossio, M. (2015). Biological autonomy: A philosophical and theoretical enquiry. Springer. https://doi.org/10.1007/978-94-017-9837-2
Moser, E. I., Kropff, E., & Moser, M.-B. (2008). Place cells, grid cells, and the brain's spatial representation system. *Annual Review of Neuroscience, 31,* 69–89. https://doi.org/10.1146/annurev.neuro.31.061307.090723
Nicolson, G. L. (2014). The fluid—mosaic model of membrane structure: Still relevant to understanding the structure, function and dynamics of biological membranes after more than 40 years. *Biochimica et Biophysica Acta — Biomembranes, 1838*(6), 1451–1466. https://doi.org/10.1016/j.bbamem.2013.10.019
Nicolson, G. L., & Ferreira de Mattos, G. (2022). Fifty years of the fluid–mosaic model of biomembrane structure and organization and its importance in biomedicine with particular emphasis on membrane lipid replacement. *Biomedicines, 10*(7), 1711. https://doi.org/10.3390/biomedicines10071711
O'Keefe, J. (2025). How the hippocampal cognitive map supports flexible navigation. *Annual Review of Neuroscience, 48*, 331–344. https://doi.org/10.1146/annurev-neuro-112723-023341
Ouyang, L., Wu, J., Jiang, X., Almeida, D., Wainwright, C. L., Mishkin, P., Zhang, C., Agarwal, S., Slama, K., Ray, A., Schulman, J., Hilton, J., Kelton, F., Miller, L., Simens, M., Askell, A., Welinder, P., Christiano, P. F., Leike, J., & Lowe, R. (2022). Training language models to follow instructions with human feedback. In *Advances in Neural Information Processing Systems 35*. Curran Associates.
Pan, A., Bhatia, K., & Steinhardt, J. (2022). *The effects of reward misspecification: Mapping and mitigating misaligned models* [Conference paper]. The Tenth International Conference on Learning Representations (ICLR 2022). https://openreview.net/forum?id=JYtwGwIL7ye
Paoli, M., Macrì, C., & Giurfa, M. (2023). A cognitive account of trace conditioning in insects. *Current Opinion in Insect Science*, 57, 101034. https://doi.org/10.1016/j.cois.2023.101034
Park, J. S., O'Brien, J. C., Cai, C. J., Morris, M. R., Liang, P., & Bernstein, M. S. (2023). Generative agents: Interactive simulacra of human behavior. *Proceedings of the 36th Annual ACM Symposium on User Interface Software and Technology*, 1–22. https://doi.org/10.1145/3586183.3606763
Pearce, J. M., & Bouton, M. E. (2001). Theories of associative learning in animals. *Annual Review of Psychology, 52,* 111–139. https://doi.org/10.1146/annurev.psych.52.1.111
Perry, C. J., Barron, A. B., & Cheng, K. (2013). Invertebrate learning and cognition: Relating phenomena to neural substrate. *WIREs Cognitive Science, 4*(5), 561–582. https://doi.org/10.1002/wcs.1248
Plotnik, J. M., de Waal, F. B. M., & Reiss, D. (2006). Self-recognition in an Asian elephant. *Proceedings of the National Academy of Sciences, 103*(45), 17053–17057. https://doi.org/10.1073/pnas.0608062103
Poort, J., Wilmes, K. A., Blot, A., Chadwick, A., Sahani, M., Clopath, C., Mrsic-Flogel, T. D., Hofer, S. B., & Khan, A. G. (2022). Learning and attention increase visual response selectivity through distinct mechanisms. *Neuron, 110*(4), 686–697, https://doi.org/10.1016/j.neuron.2021.11.016
Poulet, J. F. A., & Hedwig, B. (2007). New insights into corollary discharges mediated by identified neural pathways. *Trends in Neurosciences, 30*(1), 14–21. https://doi.org/10.1016/j.tins.2006.11.005
Raposo, D., Ritter, S., Richards, B. A., Lillicrap, T. P., Humphreys, P. C., & Santoro, A. (2024). *Mixture-of-depths: Dynamically allocating compute in transformer-based language models*. arXiv. https://doi.org/10.48550/arXiv.2404.02258
Reiss, D., & Marino, L. (2001). Mirror self-recognition in the bottlenose dolphin: A case of cognitive convergence. *Proceedings of the National Academy of Sciences, 98*(10), 5937–5942. https://doi.org/10.1073/pnas.101086398
Riley, J. R., Greggers, U., Smith, A. D., Reynolds, D. R., & Menzel, R. (2005). The flight paths of honeybees recruited by the waggle dance. *Nature, 435*(7039), 205–207. https://doi.org/10.1038/nature03526

Rivi, V., Benatti, C., Rigillo, G., & Blom, J. M. C. (2023). Invertebrates as models of learning and memory: Investigating neural and molecular mechanisms. *Journal of Experimental Biology, 226*(3), jeb244844. https://doi.org/10.1242/jeb.244844

Robert, T., Tarapata, K. & Nityananda, V. (2024). Learning modifies attention during bumblebee visual search. *Behavioral Ecology and Sociobiology*, *78*, article 22. https://doi.org/10.1007/s00265-024-03432-z

Roeder, T. (2005). Tyramine and octopamine: Ruling behavior and metabolism. *Annual Review of Entomology, 50,* 447–477. https://doi.org/10.1146/annurev.ento.50.071803.130404

Salas, C., Broglio, C., & Rodríguez, F. (2003). Evolution of forebrain and spatial cognition in vertebrates: Conservation across diversity. *Brain, Behavior and Evolution, 62*(2), 72–82. https://doi.org/10.1159/000072438

Salena, M. G., Turko, A. J., Singh, A., Pathak, A., Hughes, E., Brown, C., & Balshine, S. (2021). Understanding fish cognition: A review and appraisal of current practices. *Animal Cognition, 24*(3), 395–406. https://doi.org/10.1007/s10071-021-01488-2

Salge, C., Glackin, C., & Polani, D. (2014). Empowerment—An introduction. In M. Prokopenko (Ed.), Guided self-organization: Inception (pp. 67–114). Springer.

Sara, S. J. (2009). The locus coeruleus and noradrenergic modulation of cognition. *Nature Reviews Neuroscience, 10*(3), 211–223. https://doi.org/10.1038/nrn2573

Schubiger, M. N., Fichtel, C., & Burkart, J. M. (2020). Validity of cognitive tests for non-human animals: Pitfalls and prospects. *Frontiers in Psychology, 11*, 1835. https://doi.org/10.3389/fpsyg.2020.01835

Schuman, C. D., Kulkarni, S. R., Parsa, M., Mitchell, J. P., Date, P., & Kay, B. (2022). Opportunities for neuromorphic computing algorithms and applications. *Nature Computational Science, 2*(1), 10–19. https://doi.org/10.1038/s43588-021-00184-y

Schuster, T., Fisch, A., Gupta, J., Dehghani, M., Bahri, D., Tran, V. Q., Tay, Y., & Metzler, D. (2022). Confident adaptive language modeling. *Advances in Neural Information Processing Systems, 35*, 17456–17472. https://doi.org/10.48550/arXiv.2207.07061

Seeds, A. M., Ravbar, P., Chung, P., Hampel, S., Midgley, F. M., Jr., Mensh, B. D., & Simpson, J. H. (2014). A suppression hierarchy among competing motor programs drives sequential grooming in *Drosophila*. *eLife, 3,* e02951. https://doi.org/10.7554/eLife.02951

Seelig, J. D., & Jayaraman, V. (2015). Neural dynamics for landmark orientation and angular path integration. *Nature, 521*, 186–191. https://doi.org/10.1038/nature14446

Sherman, B. E., Turk-Browne, N. B., & Goldfarb, E. V. (2023). Multiple memory subsystems: Reconsidering memory in the mind and brain. *Perspectives on Psychological Science, 19*(1), 103–125. https://doi.org/10.1177/17456916231179146

Shi, F., Chen, X., Misra, K., Scales, N., Dohan, D., Chi, E. H., Schärli, N., & Zhou, D. (2023). Large language models can be easily distracted by irrelevant context. In *Proceedings of the 40th International Conference on Machine Learning* (Vol. 202, pp. 31210–31227). PMLR. https://proceedings.mlr.press/v202/shi23a.html

Shinn, N., Cassano, F., Berman, E., Gopinath, A., Narasimhan, K., & Yao, S. (2023). Reflexion: Language agents with verbal reinforcement learning. *Advances in Neural Information Processing Systems*, *36*, 8634–8652. https://doi.org/10.48550/arXiv.2303.11366

Silver, D., Singh, S., Precup, D., & Sutton, R.S. (2021). Reward is enough. *Artificial Intelligence, 299,* 103535. https://doi.org/10.1016/j.artint.2021.103535

Skalse, J., Howe, N. H. R., Krasheninnikov, D., & Krueger, D. (2022). Defining and characterizing reward hacking. In *Advances in Neural Information Processing Systems 35* (pp. 9460–9471). Curran Associates.

Smith, D. M., & Torregrossa, M. M. (2021). Valence encoding in the amygdala influences motivated behavior. *Behavioural Brain Research*, 411, 113370. https://doi.org/10.1016/j.bbr.2021.113370

Snell, C., Lee, J., Xu, K., & Kumar, A. (2025). Scaling LLM test-time compute optimally can be more effective than scaling parameters for reasoning. *Proceedings of the International Conference on Learning Representations (ICLR).* https://proceedings.iclr.cc/paper_files/paper/2025/hash/1b623663fd9b874366f3ce019fdfdd44-Abstract-Conference.html

Sourjik, V., & Wingreen, N. S. (2012). Responding to chemical gradients: Bacterial chemotaxis. *Current Opinion in Cell Biology, 24*(2), 262–268. https://doi.org/10.1016/j.ceb.2011.11.008

Spaethe, J., Tautz, J., & Chittka, L. (2006). Do honeybees detect colour targets using serial or parallel visual search? *Journal of Experimental Biology, 209*(6), 987–993. https://doi.org/10.1242/jeb.02124

Sterling, P. (2012). Allostasis: A model of predictive regulation. *Physiology & Behavior, 106*(1), 5–15. https://doi.org/10.1016/j.physbeh.2011.06.004

Sterling, P., & Laughlin, S. (2015). *Principles of neural design*. MIT Press.
Stone, T., Webb, B., Adden, A., Weddig, N. B., Honkanen, A., Templin, R., Wcislo, W., Scimeca, L., Warrant, E., & Heinze, S. (2017). An anatomically constrained model for path integration in the bee brain. *Current Biology, 27*(20), 3069–3085.e11. https://doi.org/10.1016/j.cub.2017.08.052
Sutton, R. S., & Barto, A. G. (2018). *Reinforcement learning: An introduction*. (2nd ed.). MIT Press.
Swaney, K. F., Huang, C. H., & Devreotes, P. N. (2010). Eukaryotic chemotaxis: A network of signaling pathways controls motility, directional sensing, and polarity. *Annual Review of Biophysics, 39,* 265–289. https://doi.org/10.1146/annurev.biophys.093008.131228
Temizer, I., Donovan, J. C., Baier, H., & Semmelhack, J. L. (2015). A visual pathway for looming-evoked escape in larval zebrafish. *Current Biology, 25*(14), 1823–1834. https://doi.org/10.1016/j.cub.2015.06.002
Tinbergen, N. (1963). On aims and methods of ethology. *Zeitschrift für Tierpsychologie, 20*(4), 410–433. https://doi.org/10.1111/j.1439-0310.1963.tb01161.x
Van Haastert, P. J. M., & Devreotes, P. N. (2004). Chemotaxis: Signalling the way forward. *Nature Reviews Molecular Cell Biology, 5*(8), 626–634. https://doi.org/10.1038/nrm1435
Vaswani, A., Shazeer, N., Parmar, N., Uszkoreit, J., Jones, L., Gomez, A. N., Kaiser, Ł., & Polosukhin, I. (2017). Attention is all you need. In *Advances in Neural Information Processing Systems 30* (pp. 5998–6008). Curran Associates.
Vogel, D., & Dussutour, A. (2016). Direct transfer of learned behaviour via cell fusion in non-neural organisms. *Proceedings of the Royal Society B: Biological Sciences, 283*(1845), 20162382. https://doi.org/10.1098/rspb.2016.2382
von Frisch, K. (1967). *The dance language and orientation of bees* (L. E. Chadwick, Trans.). Belknap Press of Harvard University Press.
Wadhams, G. H., & Armitage, J. P. (2004). Making sense of it all: Bacterial chemotaxis. *Nature Reviews Molecular Cell Biology, 5*(12), 1024–1037. https://doi.org/10.1038/nrm1524
Webb, B. (2004). Neural mechanisms for prediction: Do insects have forward models? *Trends in Neurosciences, 27*(5), 278–282. https://doi.org/10.1016/j.tins.2004.03.004
Weber, A., & Varela, F. J. (2002). Life after Kant: Natural purposes and the autopoietic foundations of biological individuality. *Phenomenology and the Cognitive Sciences, 1*(2), 97–125. https://doi.org/10.1023/A:1020368120174
Wee, C. L., Song, E., Nikitchenko, M., Herrera, K. J., Wong, S., Engert, F. & Kunes, S. (2022). Social isolation modulates appetite and avoidance behavior via a common oxytocinergic circuit in larval zebrafish. *Nature Communications*, 13, 2573. https://doi.org/10.1038/s41467-022-29765-9
Weiss, E., Liu, Y. & Wang, Q. (2026). The contribution of the locus ceruleus–norepinephrine system to the coupling between pupil-linked arousal and cortical state. *Journal of Neuroscience*, 46(3), e0898252025. https://doi.org/10.1523/JNEUROSCI.0898-25.2025
Wendelaar Bonga, S. E. (1997). The stress response in fish. *Physiological Reviews, 77*(3), 591–625. https://doi.org/10.1152/physrev.1997.77.3.591
Wheeler, J. H. R., Foster, K. R., & Durham, W. M. (2024). Individual bacterial cells can use spatial sensing of chemical gradients to direct chemotaxis on surfaces. *Nature Microbiology, 9,* 2308–2322. https://doi.org/10.1038/s41564-024-01729-3
Wiederman, S. D., & O'Carroll, D. C. (2013). Selective attention in an insect visual neuron. *Current Biology, 23*(2), 156–161. https://doi.org/10.1016/j.cub.2012.11.048
Williams, J. G. (2010). *Dictyostelium* finds new roles to model. *Genetics 185*(3), 717-726. https://doi.org/10.1534/genetics.110.119297
Yang, S., Wang, S., Si, Q., Zhou, Y., Wang, W., & Yan, Y. (2025). *Dynamic early exit in reasoning models*. arXiv. https://doi.org/10.48550/arXiv.2504.15895
Yang, J., Zhou, K., Li, Y., & Liu, Z. (2024). Generalized out-of-distribution detection: A survey. *International Journal of Computer Vision, 132*(12), 5635–5662. https://doi.org/10.1007/s11263-024-02117-4
Yao, S., Shinn, N., Razavi, P., & Narasimhan, K. (2024). *τ-bench: A benchmark for tool-agent-user interaction in real-world domains*. arXiv. https://doi.org/10.48550/arXiv.2406.12045
Yao, S., Zhao, J., Yu, D., Du, N., Shafran, I., Narasimhan, K., & Cao, Y. (2023). ReAct: Synergizing reasoning and acting in language models. *International Conference on Learning Representations (ICLR).* https://doi.org/10.48550/arXiv.2210.03629
Zador, A., Escola, S., Richards, B., Ölveczky, B., Bengio, Y., Boahen, K., Botvinick, M., Chklovskii, D., Churchland, A., Clopath, C., DiCarlo, J., Ganguli, S., Hawkins, J., Körding, K., Koulakov, A., LeCun, Y., Lillicrap, T., Marblestone, A., Olshausen, B., ... Tsao, D. (2023). Catalyzing next-generation artificial intelligence through NeuroAI. *Nature Communications, 14*(1), Article 1597. https://doi.org/10.1038/s41467-023-37180-x

Zhang, N., Guo, L., & Simpson, J. H. (2020). Spatial comparisons of mechanosensory information govern the grooming sequence in *Drosophila*. *Current Biology, 30*(6), 988–1001.e4. https://doi.org/10.1016/j.cub.2020.01.045

Zhou, K., Liu, Z., Qiao, Y., Xiang, T., & Loy, C. C. (2023). Domain generalization: A survey. *IEEE Transactions on Pattern Analysis and Machine Intelligence, 45*(4), 4396–4415. https://doi.org/10.1109/TPAMI.2022.3195549